\documentclass[review]{elsarticle}

\usepackage{hyperref,lineno}
\modulolinenumbers[5]

\journal {Medical Image Analysis}

\biboptions{authoryear}

\usepackage{epsfig,graphicx,color,amsmath,amssymb,bm,lipsum,enumitem,formatfig,amsthm}
\usepackage{subcaption}
\usepackage{float}
\usepackage{amssymb}
\usepackage{stackengine}
\usepackage{scalerel}
\usepackage{xcolor}
\usepackage{graphicx}
\newcommand\openbigstar[1][0.7]{%
  \scalerel*{%
    \stackinset{c}{-.125pt}{c}{}{\scalebox{#1}{\color{white}{$\bigstar$}}}{%
      $\bigstar$}%
  }{\bigstar}
}

\usepackage{mathrsfs} 
\usepackage{xcolor}
\usepackage{amsmath}

\usepackage{lipsum}                     
\usepackage{xargs} 
\newcommand{\ud}[1]{\textcolor{black}{#1}}
\newcommand{\revise}[1]{\textcolor{black}{#1}}
\newcommand{\revisesec}[1]{\textcolor{black}{#1}}

\providecommand {\mycite}[1]{(\cite{#1})}

\begin{document}

\begin{frontmatter}
  
  \title{Towards Lower-Dose PET using \\ Physics-Based Uncertainty-Aware Multimodal Learning with Robustness to Out-of-Distribution Data}
  
  \tnotetext[mytitlenote]{The authors V. P. Sudarshan and U. Upadhyay contributed equally.}
  
  
  \author{Viswanath P. Sudarshan$^{\openbigstar 1,2}$, Uddeshya Upadhyay$^{\openbigstar 1}$, \\ Gary F. Egan$^{3}$, Zhaolin Chen$^{3}$, Suyash P. Awate$^{1}$}
  
  
  \address{$^1$Computer Science and Engineering (CSE) Department, \\ Indian Institute of Technology (IIT) Bombay, Mumbai, India. \\ $^2$IITB-Monash
    Research Academy, \\ Indian Institute of Technology (IIT) Bombay, Mumbai, India. \\ $^3$Monash Biomedical Imaging (MBI), \\ Monash University,
    Melbourne, Australia. }

  \begin{abstract}
    {
      Radiation exposure in positron emission tomography (PET) imaging limits its usage in the studies of radiation-sensitive populations, e.g., pregnant women, children, and adults that require longitudinal imaging.
Reducing the PET radiotracer dose or acquisition time reduces photon counts, which can deteriorate image quality.
Recent deep-neural-network (DNN) based methods for image-to-image translation enable the mapping of low-quality PET images (acquired using substantially reduced dose), coupled with the associated magnetic resonance imaging (MRI) images, to high-quality PET images.
However, such DNN methods focus on applications involving test data that match the statistical characteristics of the training data very closely and give little attention to evaluating the performance of these DNNs on new {\em out-of-distribution} (OOD) acquisitions.
We propose a novel DNN formulation that models the (i)~underlying {\em sinogram-based physics} of the PET imaging system and (ii)~the {\em uncertainty} in the DNN output through the per-voxel {\em heteroscedasticity} of the residuals between the predicted and the high-quality reference images.
Our sinogram-based uncertainty-aware DNN framework, namely, suDNN, estimates a standard-dose PET image using multimodal input in the form of (i)~a low-dose/low-count PET image and (ii)~the corresponding multi-contrast MRI images, leading to improved {\em robustness} of suDNN to OOD acquisitions.
Results on in vivo simultaneous PET-MRI, \revise{and various forms of OOD data in PET-MRI}, show the benefits of suDNN over the current state of the art, quantitatively and qualitatively. 
    }
    
  \end{abstract}


\end{frontmatter}


\section{Introduction}
\label{sec:introduction}

Positron emission tomography (PET) is a molecular imaging technique that is vital in diagnosis, disease monitoring, therapy, and drug development in
various pathologies in oncology, neurology, and cardiology
\revise
{as discussed in}~\cite{ZhaolinHBM}.
The ionizing radiation involved in PET is a cause of concern in radiation-sensitive populations including pregnant women, children, and adults that
require longitudinal imaging~\mycite{vogelius2017pediatric}.
The quality of the reconstructed image depends on the number of acquired photon counts~\mycite{PETSNR}, where higher counts lead to a higher
signal-to-noise ratio (SNR).
In current applications, lowering the radioactive dose while maintaining a sufficient number of counts for acceptable image quality leads to an
increase in scan time per bed position. This can increase patient discomfort and imaging artifacts (e.g., motion-related) and reduce
scanner throughput.
Aligning with the principle of ``as low as reasonably achievable''~\mycite{ALARA}, reduced dose can potentially encourage pre-natal studies
\revisesec{(e.g.,~\cite{jones2013potential})}, early detection of brain disorders at pre-symptomatic stages \revisesec{(e.g.,~\cite{presymptomatic})}.
Furthermore, the ability to handle low-count data can enable applications in dynamic imaging regimes, e.g., functional PET imaging as shown in \revise
{ \cite{jamadar2019simultaneous,sudarshan2021incorporation,li2020analysis} } that relies on a continuous infusion of the radiotracer, where the number
of photon counts available per timeframe is substantially lower compared to conventional static PET imaging.
Hence, there is a need to achieve PET imaging at {\em low doses} or {\em low photon counts} without compromising image quality.
Thus, we propose a framework to predict a standard-dose PET image from the multimodal input in the form of (i)~a low-count PET image and (ii)~the corresponding multi-contrast \revisesec{magnetic resonance imaging (MRI)} images acquired during simultaneous PET-MRI.

Recent deep neural network (DNN) based methods for image-to-image translation enable the mapping of low-quality PET images (acquired using
substantially reduced dose), coupled with the associated MRI images, to high-quality PET images {\color{black} (e.g.,
  \cite{subtle, subtleRadiology, xiang2017deep, wang20183d}) }.
However, current DNN methods focus on applications involving test data that match the statistical characteristics of the training data closely,
and give little attention to evaluating the performance of these DNNs on new {\em out-of-distribution} (OOD) acquisitions that differ from the
distribution of images in the training set.
In the general context of PET imaging, OOD PET data could arise from several underlying factors, e.g., variations in radiotracers, anatomy, pathology, photon counts, hardware, reconstruction protocol. It is unlikely that a single learning-based model caters to all these OOD scenarios.
To deal with various OOD scenarios, for a fixed tracer and anatomical region, a good design choice is to rely on the minimum number of DNN models; e.g., this alleviates the complexity of selecting one among multiple learned models to process the data for a new subject.  
Therefore, any learned DNN model should be robust across a broad spectrum of OOD variations.
This paper focuses on the {\em robustness} to OOD data that corresponds to variation in the \revise{quality and characteristics} of the PET data.
\revise{For instance,} the number of counts available for reconstructing the PET images shows a wide variation depending on several factors as mentioned above, including the clinical/scientific application, even for a fixed tracer and anatomical region.  Typical PET scans involve photon counts (coincident events) across a wide range of $10^6$ to $10^9$ counts~\mycite{cherry2006pet} \revise {with adjusted radiotracer dose for populations under increased radiation risk, e.g., children and young adults.}
As another example, the per-voxel photon counts also depend on patient-related factors such as body-mass index (BMI) and
age~\mycite{karakatsanis2015dosage}. For subjects with higher BMI, while certain studies \revisesec{(e.g., \cite{chang2011effects,watson2005optimizing})},
suggest increasing the dose, other studies suggest increasing the scan time; while longer scans reduce patient comfort and scanner throughput, a significant increase in the dose may have undesirable side effects.
\revise{Other factors causing variations in data include:
differences in age,
differences in imaging protocols, 
subject motion, and
pathology. 
Such variations in the data  lead to \revisesec{changes} in image features such as structure, texture, contrast, and artifacts.
}

\revise{
  In the context of medical image analysis, while several works focus on improving the accuracy of the developed models, there is limited focus on addressing the uncertainty involved in interpreting the predicted outputs.  
Several works have designed DNNs to model distributions as the outputs of their intermediate and/or final layers \revisesec{ (e.g., ~\cite{srivastava2014dropout,dropout})}.
Later works have leveraged such DNN-modeling schemes for uncertainty modeling and estimation, \revisesec{(e.g., ~\cite{kendall,lakshminarayanan2017simple})}.
Recent works show that modeling uncertainties can improve the robustness of the DNN models for tasks like segmentation and regression, \revisesec{(e.g.,~\cite{jungo2018effect,wang2019aleatoric,baumgartner2019phiseg})}.
In a similar way, }
modeling the per-voxel {\em heteroscedasticity} of the residuals between predicted and reference images can improve the learned model to better adapt to the variability across real-world datasets. Exposing this heteroscedasticity as the per-voxel {\em uncertainty} in the predicted images, which allows the learned DNN to output a distribution of PET images, may potentially aid in clinical interpretation~\mycite{nair2020exploring,wang2019aleatoric}.
DNNs that do not inform about the predicted images' underlying risk can lead to misleading outputs, especially when presented with OOD data. Thus, we propose a modeling and learning strategy that is aware of this uncertainty in the predicted outputs.
Several DNN learning methods show the benefits of transform-domain loss functions during learning, {\color{black} where the transform domain refers to a
manifold or feature space obtained from transforming the images (both predicted and ground truth). 
An example of transform-domain loss is the k-space-domain loss employed for undersampled MRI image reconstruction~\mycite{DAGAN}.}  In the same way, we propose a {\em transform-domain loss} that is motivated by the physics of the image acquisition process, where the transform domain is the {\em sinogram} domain for PET imaging.
  Our sinogram-based uncertainty-aware DNN, namely, suDNN, framework predicts a standard-dose PET (SD-PET) image from the multimodal input in the form of (i)~a low-count PET image (being low quality) and (ii)~the corresponding multi-contrast MRI images, leading to improved robustness of the learned DNN model to OOD acquisitions.
  By designing the DNN input as a combination of the low-count PET image and the multi-contrast MRI, our framework leverages aspects of learning relating to both image quality enhancement and inter-modality image-to-image translation.

This paper makes several contributions.
We propose a DNN framework to predict a standard-dose PET image from the multimodal input in the form of (i)~a low-count PET image and (ii)~the
corresponding multi-contrast MRI images acquired during simultaneous PET-MRI.
We propose a novel DNN formulation that models (i)~the underlying sinogram-based physics of the PET imaging system and (ii)~the uncertainty in the
predicted output through the per-voxel heteroscedasticity of the residuals between predicted and reference images.
\revise{
  Compared to the current state of the art, our sinogram-based uncertainty-aware DNN framework, namely, suDNN, leads to improved robustness to
  OOD acquisitions as shown by quantitative and qualitative evaluations on {\em in vivo} simultaneous PET-MRI.  This paper focuses on PET-MRI of the human brain using the [18-F] fluorodeoxyglucose (FDG) radiotracer.
}

\section{Related Work}
\label{sec:relatedWork}

Current systems for simultaneous PET-MRI typically require long acquisition times (around 20 minutes) for multi-contrast MRI scans, thereby leading to
PET scans of equivalent duration.  The long acquisition time enables lower-dose PET imaging in comparison to typical systems acquiring PET and X-ray
computed tomography (PET-CT)~\mycite{karakatsanis2015dosage}.
However, with an increasing emphasis on reducing the acquisition time in MRI, within simultaneous PET-MRI, e.g, works in ~\cite{Ehrhardt2014,sudarshan2020joint,miccai2019}, it
is important to enable PET imaging with reduced scan times or reduced photon counts (per bed position). 
Prior works on PET image enhancement can be classified into: (i)~regularized reconstruction techniques from acquired PET data and
(ii)~post-reconstruction techniques without and with learning-based approaches.
 
{\bf Regularized PET reconstruction methods:}
These refer to modeling prior knowledge, e.g., using total variation (TV) as in ~\cite{EMTV} or anatomical information from the co-registered MRI image
within the PET reconstruction routine as in~\cite{Leahy1991,Nuyts2007,Georg2018}.
Recently, \cite{MICCAI2018,sudarshan2020joint} showed that a joint dictionary model for both PET and MRI images shows improved robustness to noise-level
perturbations in the PET images.  However, that work focused on improving the noisy PET images and {\em not} on the reduction of radiotracer dosage
levels.
\cite{Kim2018} employ a DNN to improve PET image quality within the iterative PET reconstruction framework to achieve a dose reduction factor (DRF) of
around $6 \times$.
However, in many use cases, the raw list-mode data is either unavailable or entails complex mathematical models for accurately modeling the details of
the scanner physics and measurement errors. Hence, there is interest in post-reconstruction methods for image quality enhancement.

{\bf Post-reconstruction image quality enhancement without learning-based models.}
Most commonly, this involves Gaussian filtering the reconstructed PET image.
Improvements over the post-reconstruction Gaussian-smoothing approach come from methods that use higher-order statistical models for the PET image
\revisesec{(e.g.,~\cite{bagci2013,dutta2013non})}. Improvements also come from joint modeling of dependencies across co-registered PET and MRI images
\revise{as shown in}~\cite{Song2019PET}.
Recently, \cite{conditionalDIP} propose an unsupervised model for PET image denoising by employing a conditional deep image prior (DIP) that uses the
subject's anatomical MRI or CT as the input to the DNN mapping.  These methods focus on denoising, instead of dealing with smaller radiation doses.

{\bf Post-reconstruction image quality enhancement with learning-based models.}
\revise{
  Recent works like~\cite{upadhyay2019robust,masutani2020deep,qiu2020super,upadhyay2019mixed} have shown successful application of DNN based methods for image-quality enhancement. In the context of PET quality enhancement, 
}
some early works~\cite{dinggangRegressionForest,dinggangDictionary} show that learning-based approaches, e.g., regression forests and sparse
dictionary modeling, can synthesize SD-PET images from LD-PET images at a DRF of around $4 \times$.
For a similar DRF,
(i)~\cite{xiang2017deep} propose a DNN that uses an auto-context strategy to estimate patches in the SD-PET image based on the patches in the input
set of LD-PET and T1w MRI images and
(ii)~\cite{wang20183d} employ a generative adversarial network (GAN) framework, where the input to the generator is a fused version of the
multi-contrast MRI images and the LD-PET image.
\cite{resnetTRPMS} uses a ResNet architecture to learn a mapping from the noisy LD-PET image to the SD-PET image (without any dose reduction), where
the training includes a VGG-based loss term.
Recent pioneering work by~\cite{subtle} shows that it is possible to achieve a DRF of around $200 \times$ using a DNN to map the residual between
the LD-PET image and the reference SD-PET image, where the DNN uses an 2.5D-style input to mimic volumetric mapping using a lighter and computationally
cheaper model.
Subsequently, \cite{subtleRadiology} shows that, with a similar architecture and training strategy, including MRI images as input produced better
image quality than using PET images alone.  A slightly different strategy by \cite{sanaat2020projection} shows that learning a mapping between the
LD-PET sinogram and the SD-PET sinogram can lead to some improvement in the reconstructed SD-PET images, compared to the strategy of learning the
mapping from LD-PET to SD-PET in the spatial image domain. However, as mentioned earlier, the measured raw sinogram data might be either unavailable
or lead to complex models for direct integration into the DNN framework.  On the other hand, linear models of the scanner-specific sinogram
transformations are readily available, constructed using the knowledge of scanner geometry \revisesec{(e.g., ~\cite{GATE})}. The retrospectively estimated sinogram data
can model reasonably well the acquired sinogram data obtained after typical error-correction steps applied to the PET raw data.

The prior works discussed in this section employ loss functions either exclusively in the spatial domain or exclusively in the sinogram domain, but
{\em not} both. Several DNN-based methods for undersampled MRI reconstruction have shown that including a transform-domain (k-space) loss function in
addition to the spatial-domain loss function can improve the quality of reconstructed images at higher undersampling
levels \revisesec{(e.g., ~\cite{schlemper2017deep} and ~\cite{DAGAN})}.
Second, the prior works, including those for PET and MRI reconstruction, do {\em not} evaluate the models for robustness to OOD data in new
acquisitions, which are essential for clinical translation.
Third, typical PET reconstruction methods seldom quantify the uncertainty in the DNN output. Modeling uncertainty in DNNs can potentially (i)~inform
the radiologist about the imperfections in reconstructions, which may aid in clinical decision making or subsequent automated post-processing of
reconstructed images, and (ii)~provide improved performance when the DNN is presented with OOD data.  
\revisesec{
Early influential work in ~\cite{hinton2012improving} showed that the performance of DNNs can be improved by employing dropout-based regularization to reduce the problem of co-adaptation during learning. Subsequently, work in ~\cite{dropout,gal2017concrete} provided a Bayesian interpretation of dropouts within a variational learning framework and used it to estimate model-related uncertainty. In the works by ~\cite{hinton2012improving,srivastava2014dropout}
the dropout probability was a tunable free parameter. On the other hand, in the later works by 
~\cite{dropout,gal2017concrete} , the dropout probability parameter was a hidden variable within the learning framework. More recent work in ~\cite{kendall} improved uncertainty model that quantifies both model-related and data-related uncertainty.
}
The uncertainty-related works discussed above propose to estimate the uncertainty in the outputs, during training and testing phases, using stochastic layers in the DNN architecture.
In the context of medical image analysis, recent works like~\cite{jungo2018effect,wang2019aleatoric,jungo2019assessing,baumgartner2019phiseg,nair2020exploring} discuss the uncertainty estimation for medical image segmentation, and other works like~\cite{tanno2021uncertainty,sentker2018gdl,armanious2021age} discuss uncertainty estimation for various medical image regression tasks such as image enhancement for diffusion MRI, image registration, and biological age estimation using MRI, respectively.  

\revise { Our novel DNN framework, }
(i)~leverages the underlying physics of the PET imaging system and (ii)~models the uncertainty in the DNN output through the per-voxel
heteroscedasticity of the residuals between the predicted and the high-quality reference images.
Our results on a cohort of 28 subjects with {\em in vivo} PET-MRI acquisition demonstrate (i)~improved quality of the reconstructed images and
(ii)~improved robustness of the learned model in reconstructing OOD PET data as compared to state-of-the-art methods.  
\revise{ Additionally, 
compared to state of the art, we show
  that our proposed model is robust to OOD data arising from other factors such as 
  differences in imaging protocol on another cohort, motion artifacts, age, pathology, inter-scanner variability, as detailed in
  Section~\ref{sec:results}.  
  }

\section {Methods}
\label{sec:methods}

We describe suDNN's mathematical formulation, architecture, and the training strategy, for estimating SD-PET images using the multimodal input data.

\begin{figure}
  \centerline{\includegraphics[width=\columnwidth]{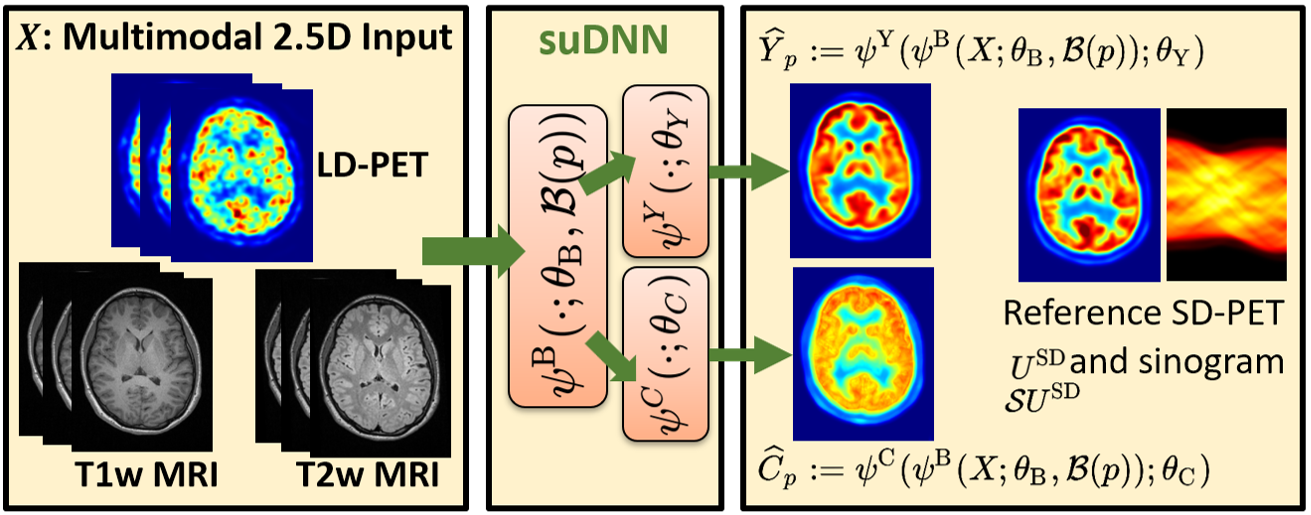}}
  \caption
      {
        {\bf Proposed suDNN Framework}.
        The inputs $X$ to suDNN are:
        (i)~the low-dose/count PET image $U^{\text{LD}}$ and
        (ii)~the multicontrast MRI images $V^{\text{T1}}$ and $V^{\text{T2}}$, incorporating the 2.5D-style training scheme.
        \revise
        {
        The suDNN models the mapping
        $
        \Psi (\cdot; \theta_{\text{B}}, \theta_{\text{Y}}, \theta_{\text{C}}, \mathcal{B}(p))
        :=
        (
        \psi^{\text{Y}} (\psi^{\text{B}} (\cdot; \theta_{\text{B}}, \mathcal{B}(p)); \theta_{\text{Y}}),
        \psi^{C}       (\psi^{\text{B}} (\cdot; \theta_{\text{B}}, \mathcal{B}(p)); \theta_{\text{C}})
        )$, where
        $\psi^{\text{B}} (\cdot; \theta_{\text{B}}, \mathcal{B}(p))$ 
        denotes a common backbone feeding into two disjoint heads
        $\psi^{\text{Y}} (\psi^{\text{B}} (\cdot; \theta_{\text{B}}, \mathcal{B}(p)); \theta_{\text{Y}})$ and
        $\psi^{C}       (\psi^{\text{B}} (\cdot; \theta_{\text{B}},  \mathcal{B}(p)); \theta_{\text{C}})$.
        The $\theta_{\cdot}$ variables denote the parameters of each component.
        $\mathcal{B}(p)$ is a Bernoulli random variable, with parameter $p$, modeling the dropout.
        }
        The suDNN outputs are:
        (i)~the high-quality PET image modeled by random field $\widehat{Y}$ and
        (ii)~the random field $\widehat{C}$ modeling the per-voxel variances in the residuals between the predicted image and the reference SD-PET
        image $U^{\text{SD}}$.
      }
      \label{fig:model}
\end{figure}

\subsection{suDNN Modeling}
\label{sec:model}

Let random fields $U^{\text{LD}}$ and $U^{\text{SD}}$ model the acquired LD-PET and SD-PET images, respectively, across the population.
Let random fields $V^{\text{T1}}$ and {\color{black} $V^{\text{T2}}$} model the acquired T1w and T2w MRI images, respectively.
For each subject, the PET and MRI images ($U^{\text{LD}}$, $V^{\text{T1}}$, $V^{\text{T2}}$, and $U^{\text{SD}}$) are spatially co-registered to a
common coordinate frame, where each image contains $K$ voxels.
We propose to learn the suDNN by relying on a multimodal image-to-image translation framework \revise { incorporating a dropout-based statistical
model, for improved regularization during learning, involving a Bernoulli random variable $\mathcal{B}(p)$ with parameter $p$ as described in
~\cite{tompson2015efficient}}. Thus, our framework takes as input the random-field triplet
$X := \{ U^{\text{LD}}, V^{\text{T1}}, V^{\text{T2}} \}$
\revise { and maps it to output (i)~a distribution on the possible SD-PET images associated with the input $X$, along with 
(ii)~a distribution on the possible per-voxel variances of the (heteroscedastic) residuals associated with the predicted SD-PET images, the square root of which can also be interpreted as the per-voxel uncertainties associated with the predicted SD-PET images. }
Thus, the suDNN models a \revise{stochastic} regressor $\Psi (\cdot; \Theta, \mathcal{B}(p))$, parameterized by weights $\Theta$ and the 
dropout-probability parameter $p$, such that
$\Psi (X; \Theta, \mathcal{B}(p)) := (\widehat{Y}_p, \widehat{C}_p)$, \revise{where $\widehat{Y}_p$ and $\widehat{C}_p$ characterize distributions on the SD-PET images and on
their associated per-voxel uncertainties, respectively. 
In this way, $\widehat{Y}_p$ and $\widehat{C}_p$ are also stochastic outputs where the stochasticity stems from
the underlying dropout layer involving parameter $p$, as detailed in the next paragraph.}
suDNN learns the regressor using the training set $\mathcal{T} := \{ X_i \cup U_i^{\text{SD}} \}_{i=1}^N$ comprising images from $N$ subjects.
Figure~\ref{fig:model} shows our suDNN framework.

We propose a DNN model that is based on a U-Net architecture~\mycite{ronneberger2015u}.
The proposed suDNN differs from the standard U-Net by incorporating:
(i)~multimodal input where the data from the PET, T1w MRI, and T2w MRI images are treated as different channels,
(ii)~a 2.5D-style (similar to the strategy in~\cite{subtleRadiology}) where the estimation of a particular slice in the SD-PET image takes as input,
from each modality, a collection of slices in the neighborhood,
(iii)~a dual-head output (Figure~\ref{fig:model}), where the output from one DNN head represents the predicted SD-PET images, and the output from the
other head represents the per-voxel variances modeling the variability in the predicted SD-PET images, inspired by~\mycite{kendall}, and
(iv)~a dropout model~\mycite{srivastava2014dropout}, following its bottleneck layer, for regularization during learning.
Specifically, suDNN models the mapping
\revise{
\begin{linenomath*}
\begin{align}
  \Psi (\cdot; \theta_{\text{B}}, \theta_{\text{Y}}, \theta_{\text{C}}, \mathcal{B}(p))
  :=
  (
  \psi^{\text{Y}} (\psi^{\text{B}} (\cdot; \theta_{\text{B}}, \mathcal{B}(p)); \theta_{\text{Y}}),
  \psi^{C}       (\psi^{\text{B}} (\cdot; \theta_{\text{B}}, \mathcal{B}(p)); \theta_{\text{C}})
  )
  ,
\end{align}
\end{linenomath*}
} where a single convolutional backbone represented by $\psi^{\text{B}} (\cdot; \theta_{\text{B}}, \mathcal{B}(p))$,
parameterized by $\theta_B$ and the Bernoulli random variable $\mathcal{B}$ parameterized by $p$,
feeds the resulting latent features to the two {\em disjoint} output heads, i.e., one for representing the predicted images denoted by the mapping
$\psi^{\text{Y}} (\cdot; \theta_{\text{Y}})$ and the other for representing the variance images denoted by the mapping $\psi^{\text{C}}(\cdot;
\theta_{\text{C}})$.
Thus, for a given multimodal input $X$
and the set of parameters $\Theta := \theta_{\text{B}} \cup \theta_{\text{Y}} \cup \theta_{\text{C}}$, 
the suDNN outputs \revise{
  $\widehat{Y}_p := \psi^{\text{Y}} (\psi^{\text{B}} (X; \theta_\text{B}, \mathcal{B}(p)); \theta_\text{Y})$ and
  $\widehat{C}_p := \psi^{\text{C}} (\psi^{\text{B}} (X; \theta_\text{B}, \mathcal{B}(p)); \theta_\text{C})$.
}

\subsection {Uncertainty-Aware and Physics-Based Loss Functions}
\label{sec:loss_function}

A mean squared error (MSE) loss function between the DNN-output PET image $\revise{\widehat{Y}_p}$ and the high-quality PET image $U^{\text{SD}}$
assumes homoscedasticity of the per-voxel residuals, which may turn out to be a gross approximation in general, and especially so in the context of
OOD data.
Thus, we propose a loss function that explicitly adapts to the heteroscedasticity of the per-voxel residuals between the output PET image
$\revise{\widehat{Y}_p}$ and the high-quality PET image $U^{\text{SD}}$.  Our empirical evaluation (later) shows that such a model leads to the
robustness of the learned model to OOD PET test data.
Thus, for each subject, we model the output of suDNN as a pair consisting of (i)~the predicted SD-PET images $\revise{\widehat{Y}_p}$ and (ii)~the
images $\revise{\widehat{C}_p}$ modeling the per-voxel variances in the residuals between the predicted images and the reference SD-PET image.
An alternate interpretation for the values in $\revise{\widehat{Y}_p}$ and $\revise{\widehat{C}_p}$ stems from the notion of a DNN that outputs a
family of images modeled by a Gaussian distribution, where $\revise{\widehat{Y}_p}$ models the per-voxel means and $\revise{\widehat{C}_p}$ model the
per-voxel variances.
We find that incorporating this uncertainty-aware (or heteroscedasticity-based) loss leads to improved robustness to OOD acquisitions.
\revise { Thus, }
we propose loss functions that enforce similarity in two domains, i.e., (i)~the spatial domain and (ii)~the sinogram domain modeling the PET detector geometry.
\revise{
  We find that incorporating the transform-domain (sinogram-domain) loss and modeling the per-voxel heteroscedasticity in both domains make our model robust to OOD acquisitions.
}
The overall loss function of the suDNN, $\mathcal{L}_{\text{SU}}$, is a weighted combination of two loss functions, i.e.,
(i)~uncertainty-aware loss in the image space $\mathcal{L}_{\text{U}}$ and
(ii)~uncertainty-aware PET-physics-based loss in the sinogram space $\mathcal{L}_{\text{S}}$.

{\bf Uncertainty-Aware Spatial-Domain Loss $\mathcal{L}_\text{U}$.}
For input image $X_i$, let $\revise{\widehat{Y}_{pi}[k]}$ represent the $k$-th voxel in the spatial domain for the $i$-th predicted image
$\revise{\widehat{Y}_{pi}}$, and let $\revise{\widehat{C}_{pi}[k]}$ represent the $k$-th voxel for the $i$-th predicted variance image
$\revise{\widehat{C}_{pi}}$.  We employ a Gaussian likelihood model for the observed image $U^{\text{SD}}_i$ in the image space, parameterized by
per-voxel means in \revise{$\widehat{Y}_{pi} = \psi^{\text{Y}} (\psi^{\text{B}} (X_i; \theta_\text{B}, \mathcal{B}(p)); \theta_\text{Y})$} and
per-voxel variances in \revise{$\widehat{C}_{pi} = \psi^{\text{C}} (\psi^{\text{B}} (X_i; \theta_\text{B}, \mathcal{B}(p)); \theta_\text{C})$}.
Thus, the negative of the log-likelihood function leads to the loss over the training-set $\mathcal{T}$ as
\begin{linenomath*}
  \begin{align}
    \mathcal{L}_\text{U} (\Theta; \mathcal{T})
    :=
    \sum_{i=1}^{N}
    \sum_{k=1}^{K}
    {\revise{\mathbb{E}_{P_{\mathcal{B}(p)}} }}
    {\revise{\bigg [ }}
    \frac
        {    ( {\revise{\widehat{Y}_{pi}[k]}} - U^{\text{SD}}_{i}[k] )^2 }
        {      {\revise{\widehat{C}_{pi}[k]}} + \epsilon            }
        +
        \log ( {\revise{\widehat{C}_{pi}[k]}} + \epsilon)
    {\revise{\bigg ]}}
    ,
  \end{align}
  \label{eqn:image_space}
\end{linenomath*}
where $\epsilon \in \mathbb{R}^+$ is a small constant for numerical stability.
\revise { Here, $N$ denotes the number of training samples, $K$ the number of voxels in each image, and $\mathbb{E}_{P_{\mathcal{B}(p)}}$ represents
  expectation under the Bernoulli probability distribution characterizing $\mathcal{B}(p)$. The above method of modeling uncertainty in the spatial
  domain is similar to \cite{kendall}.  }
Equation~\ref{eqn:image_space} consists of two components:
(i)~the per-voxel squared residual/error (\revise{$\widehat{Y}_{pi}[k]$}$- U^{\text{SD}}_i[k])^2$ scaled down by the variance
\revise{$\widehat{C}_{pi}[k]$}, and
(ii)~the penalty term $\log (\revise{\widehat{C}_{pi}[k]} + \epsilon)$ on the per-voxel variance \revise{$\widehat{C}_{pi}[k]$}, which penalizes large
values of \revise{$\widehat{C}_{pi}[k]$}.
We enforce positivity on the elements of the suDNN outputs $\revise{\widehat{Y}_{pi}}$ using ReLU activation function in the final layer of the head
modeling $\psi^{Y}$.
We enforce positivity of $\revise{\widehat{C}_{pi}}$ by employing an exponentiation layer as the final layer of $\psi^{C}$.
suDNN learning does {\em not} require explicit supervision in the form of ground-truth observations for $\revise{\widehat{C}_{pi}}$, but rather learns
to map to $\revise{\widehat{C}_{pi}}$ using the loss in Equation~\ref{eqn:image_space} using the SD-PET image data $U^{\text{SD}}$.

{\bf Uncertainty-aware Sinogram-Domain Loss $\mathcal{L}_\text{S}$.}
Let operator $\mathcal{S}$ model the linear sinogram transformation associated with PET image acquisition for each transaxial slice.  The operator
$\mathcal{S}$ takes a 2D image with $K$ voxels and produces a sinogram with $L$ discrete elements.  Because we model the per-voxel residual
($\revise{\widehat{Y}_{pi}} - {U}^\text{SD}_i$) in the spatial domain by a Gaussian distribution, the per-element residual in the sinogram domain also
follows a Gaussian distribution.  Similarly, given that \revise{$\widehat{C}_{pi}$} models the heteroscedasticity of the Gaussian-distributed
residuals across the voxels in the spatial domain, we propose to model the distribution of the residuals in the sinogram domain as a factored
multivariate Gaussian (one factor per element), with the per-element variances of the sinogram-domain residual $\mathcal{S} \revise{\widehat{Y}_{pi}}
- \mathcal{S} {U}^\text{SD}_i$ being $\mathcal{S} \revise{\widehat{C}_{pi}}$.  For simplicity, we exclude modeling the covariances between the
per-voxel residuals in the sinogram domain resulting from the dependencies introduced by the sinogram operator $\mathcal{S}$.  Thus, we propose a
physics-based loss term in the sinogram domain as
\begin{linenomath*}
\begin{align}
  \mathcal{L}_\text{S} (\Theta; \mathcal{T})
  :=
  \sum_{i=1}^{N}
  \sum_{l=1}^{L}
  {\revise{\mathbb{E}_{P_{\mathcal{B}(p)}} }}
  {\revise{\bigg [}}
  \frac
      {    ( \mathcal{S} {\revise{\widehat{Y}_{pi}}}[l] - \mathcal{S} U^{\text{SD}}_i[l] )^2 }
      {      \mathcal{S} {\revise{\widehat{C}_{pi}}}[l] + \tau }
      +
      \log ( \mathcal{S} {\revise{\widehat{C}_{pi}}}[l] + \tau )
  {\revise{\bigg ]}}
      ,
\end{align}
\end{linenomath*}
where $\tau \in \mathbb{R}^+$ is a small constant for numerical stability.

{\bf Overall Loss Function $\mathcal{L}_\text{SU}$.}
We propose to optimize the set of parameters $\Theta$ of our DNN by minimizing the overall loss function consisting of uncertainty-aware loss
functions in both the image-space and the sinogram-space given by
\begin{linenomath*}
  \begin{align}
    \mathcal{L}_\text{SU} (\Theta; \mathcal{T})
    :=
    \mathcal{L}_\text{U}  (\Theta; \mathcal{T})
    +
    \lambda
    \mathcal{L}_\text{S}  (\Theta; \mathcal{T})
    ,
  \end{align}
\end{linenomath*}
where $\lambda$ is a non-negative real-valued free parameter that controls the weight of the physics-based sinogram-domain loss.
In this work, we tune the value of $\lambda$ using a validation set. 

\subsection{DNN architecture and learning strategy}

\begin{figure}
  \centerline { \includegraphics[width=0.72\columnwidth] {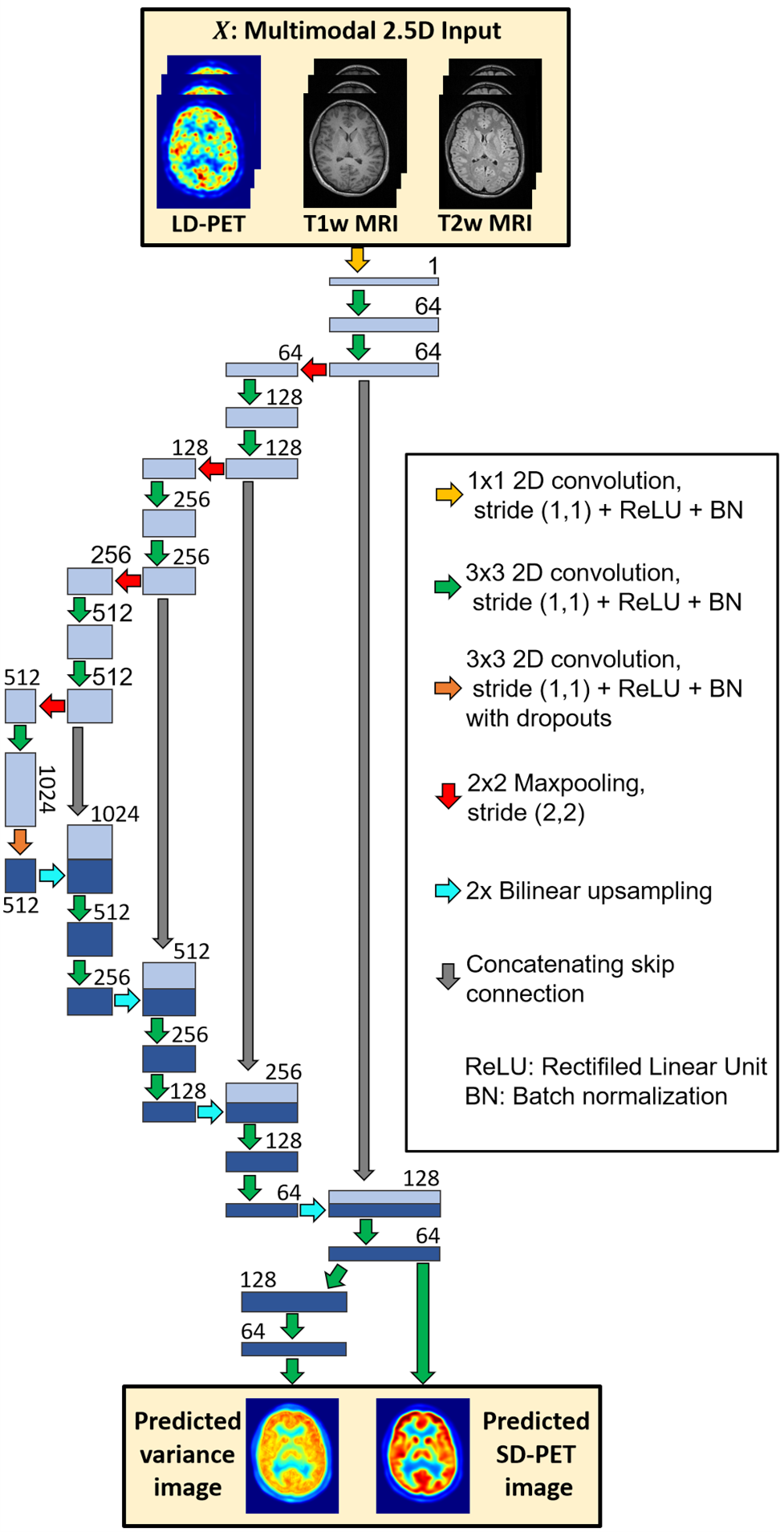} }
  \caption
      {
        {\bf suDNN Architectural Details.}
        The numbers adjoining the blue boxes indicate the number of feature maps obtained at that stage.
      }
      \label{fig:architecture}
\end{figure}

Figure~\ref{fig:architecture} shows the details of the suDNN architecture.
We employ a U-Net architecture comprising an encoder and a decoder that have a symmetric structure, and incorporate skip connections from the encoder
to the decoder. Both the encoder and decoder comprise three convolutional blocks. The downsampling/upsampling layers downsample/upsample by a factor
of two.
After every convolutional layer, suDNN uses standard batch \ud{normalization~\mycite{ioffe2015batch}} and ReLU activation.
The bottleneck layer is followed by a dropout layer (\revise{characterized by $\mathcal{B}(p)$}) for
\ud{regularization~\mycite{srivastava2014dropout}}, using a dropout-probability value of \revise{$p = 1/1024$} during training as well as inference.
\revise{Here, $p$ is a hyperparameter, set such that it drops on an average one channel (out of 1024 channels, see Figure~\ref{fig:architecture}) at
  the bottleneck layer per forward pass.  }
suDNN uses the Adam optimizer~\mycite{adam} during training, including $\ell_2$ regularization on the weights, for 500 epochs, with an initial
learning rate of $\gamma = 0.00003$.
suDNN employs a cosine annealing strategy for updating $\gamma$.
\revise { During inference, we rely on the dropout layer to generate the multiple outputs for a given input $X_i$ by performing multiple forward
 passes, say $M$ (here, $M = 50$), through the DNN with dropouts activated, yielding a set of outputs 
 $\{ \widehat{Y}^{m}_{i}, \widehat{C}^{m}_{i}\}_{m=1}^M$.
 Here,  $\widehat{Y}^{m}_{i}$ is a particular sampled instance of the stochastic output $\widehat{Y}_{pi}$.
We infer the final predicted images as the averages of the $M$ outputs, i.e., 
$\widehat{Y}_i :=  ( {1}/{M}) \sum_{m=1}^M \widehat{Y}^m_{i} $ and $\widehat{C}_i := ({1}/{M}) \sum_{m=1}^M \widehat{C}^m_{i} $. }

\section{Experiments and Results}
\label{sec:results}

This section describes the {\em in vivo} data acquired for this work, the baseline methods used for comparison, the empirical analyses for evaluating
the robustness of all methods to OOD degradations in the input data, and ablation studies to analyze the contribution of various components in the
suDNN framework.

\subsection {In vivo Data}
\label{sec:data}

We acquired data using simultaneous PET-MRI in a cohort of 28 healthy individuals \revise{ (volunteers with mean age 19.6 years and standard deviation 1.7 years, including 21
  females)} on a 3T Siemens Biograph mMR system, following institute ethics approval.  The average dose administered for each subject was
approximately 230 MBq {\color{black}F-18-FDG}.  The MRI contrast images, i.e., ultra-short echo time (UTE), T1 MPRAGE, and T2-SPC, were acquired during
the PET scan.
The SD-PET image was reconstructed using counts obtained over a duration of 30 minutes, starting 55 minutes after the administration of the tracer.
The total number of useful counts over the 30-minute duration used for reconstruction of the SD-PET image were around $600 \times 10^6$. To simulate
the LD-PET data, we randomly selected around $3.4 \times 10^6$ counts, spread uniformly over the scan duration, resulting in a DRF of around $180
\times$.
For attenuation correction, pseudo-CT maps generated using the UTE images~\mycite{Burgos2014} were employed.
Both the SD-PET and LD-PET images were reconstructed using proprietary software using ordinary-Poisson ordered-subset expectation-maximization
(OP-OSEM) algorithm with three iterations and 21 subsets, along with point spread function (PSF) modeling and a post-reconstruction Gaussian
smoothing.  The software produced reconstructed PET images of voxel sizes 2.09 $\times$ 2.09 $\times$ 2.03~mm$^3$.  The voxel size for the
reconstructed MRI images was 1~mm$^3$ isotropic.
For each subject, the LD-PET, SD-PET, and the T2w MRI images were registered (using rigid spatial 
\revise
{
transformation}) 
and resampled to the T1w MRI image
space using ANTS~\mycite{ANTS} software.
For the task of predicting SD-PET images from the input set of LD-PET, T1w MRI, and T2w MRI images, we randomly selected 20 subjects for training, 2
subjects for validation, and the remaining for testing.  For each subject, we obtained 100 transaxial slices (around 70 slices within the cerebrum and
around 30 slices in the cerebellum).

\subsection {Baseline Methods}

We evaluate the performance of the proposed suDNN in comparison to {\em five} recently proposed DNN-based methods for SD-PET prediction.  
For a fair
comparison, we incorporate a 2.5D-style (similar to the strategy in~\cite{subtleRadiology}) training scheme for all other methods.  
\revise
{
That is, to produce a predicted image for a given slice,
we use five slices as the input of the DNN (one central, two above, and two below).
}
The baseline methods are as follows.
\begin{itemize}
\item %
  {\bf M1: Conditional DIP.}  M1 is an unsupervised method based on conditional DIP in~\mycite{conditionalDIP}. The method is unsupervised and does
  {\em not} rely on any training data.  As proposed in \cite{conditionalDIP}, the input to the DNN is the structural MRI image.  
  We use a U-Net as in ~\cite{ronneberger2015u} modified to accept a two-channel input (T1w and T2w MRI).
  For this method, we
  use the validation set to tune the optimal number of epochs, to maximize the SSIM between the predicted PET image and the reference SD-PET image.
\item %
  {\bf M2: Unimodal ResNet with perceptual loss.}  M2 is similar to the framework proposed in~\mycite{resnetTRPMS}. 
  M2 uses only the PET image (unimodal) as input, with a standard ResNet architecture~\mycite{resnetTRPMS}, and employs a perceptual loss that is based on features obtained from a VGG network trained on natural images.
\item %
  {\bf M3 and M4: 2.5D unimodal and multimodal U-Net, respectively.}  Both M3 and M4 use the architecture described in~\mycite{subtle}. M3 uses only
  the PET image as (unimodal) input~\mycite{subtle}. M4 uses PET and multi-contrast MRI images as multi-channel input~\mycite{subtleRadiology}.  Both
  M3 and M4 explicitly model and estimate the residuals between the input LD-PET and the reference SD-PET image.
\item %
  {\bf M5: Multi-channel GAN.}  M5 is similar to the GAN-based model in~\mycite{wang20183d} that uses multi-channel input comprising PET and
  multi-contrast MRI images, including diffusion-weighted MRI. Because of the unavailability of diffusion-weighted MRI images for our dataset, and for
  a fair comparison with all the other methods, we use only the T1w and T2w MRI images for training. The model in~\mycite{wang20183d} employs a   anatomical-region-specific learnable 1$\times$1 convolution layer to produce a fused image that becomes the input to the generator of the GAN.
  We employ a 2.5D U-Net-based architecture for the generator.
\end{itemize}
{
\color{black}
M1 and M2 focus on denoising and {\em not} on dose reduction. 
}
M3--M5 propose to achieve DRFs in the range 4--200.
\revise { DNNs M1, M3, M4, M5, and suDNN employ similar U-Net-based backbone architecture with comparable number of parameters.  On the other hand, M2
  employs a ResNet as described above, with significantly more parameters compared to other DNNs.  For all the DNNs that necessitate a
  training stage (M2--M5 and suDNN), we use the same training-validation-testing split.  The hyperparameters for all the DNNs are tuned using the
  validation set.  We trained all the DNNs with a decaying learning rate for \revisesec{500} epochs.  In practice, we observed that all the models converged
  within 300--400 epochs.  For each DNN, we selected the model that provided the best performance (SSIM) on the validation set.  } For quantitative
evaluation of the quality of the predicted PET image, with respect to the reference SD-PET image, we use (i)~peak SNR (PSNR) and (ii)~structural
similarity index (SSIM)~\mycite{SSIM}.

{\color{black}
\subsection{Out-of-distribution (OOD) data}
\label{sec:robustness}

For training all the DNNs discussed in this paper, we use the training set of LD-PET images from a single cohort discussed in Section~\ref{sec:data}.
Primarily, we evaluate the performance of all the methods on the testing set of LD-PET images.  In a practical setting, even with a fixed scanner and imaging protocol (i.e., the acquisition schemes for MRI contrasts and the radiotracer used for PET), various factors are contributing to OOD
data, e.g., variation in photon-count statistics due to slight variations in the injected dose, physiological factors like body mass index (BMI),
aging brain, pathology.
In addition to the above, several other factors can contribute to OOD data, as described in Section~\ref{sec:introduction}.
Hence, to evaluate the generalizability of the proposed model, we provide a comprehensive evaluation on OOD data arising from several acquisition scenarios:
(i)~variation due to reduced photon counts (reduced SNR), 
(ii)~variation due to patient motion,
(iii)~variation due to pathology (Alzheimer's disease) and age, and
(iv)~variation due to PET and MRI data acquired using separate scanners or different imaging protocols.
We now discuss the above-mentioned OOD datasets in detail.
}

\textbf{OOD data with variation in photon counts or SNR (OOD-Counts).}  We generate OOD PET data by varying the photon counts and the associated SNR
in the sinogram space, followed by OSEM reconstruction with post-reconstruction Gaussian smoothing.  We generate two additional sets of test data at
increasing degradation levels in the input LD-PET data, namely very low-dose (vLD-PET) and ultra-low-dose (uLD-PET).  We generate the OOD test set
consisting of vLD-PET and uLD-PET as follows.

We retrospectively (i)~scale down the intensities in the LD-PET image, (ii)~forward-project the resulting scaled-down LD-PET image using the sinogram
operator $\mathcal{S}$, (iii)~introduce Poisson noise in the sinogram domain on the projected image, and (iv)~perform OSEM-based reconstruction to get
the input vLD-PET or uLD-PET image.
For forward projection of the LD-PET images, we use the projection model from STIR~\mycite{thielemans2012stir} that is based on a ray-tracing
algorithm for the system geometry, which is similar to that used in the Siemens PET-MRI system used in this study.
The PSNR value, averaged across the test set, between the reference SD-PET image and LD-PET image was around 21 dB.  To obtain vLD-PET and uLD-PET, we
scale the LD-PET images such that, after OSEM reconstruction, the PSNR values, averaged across the test set, between the reference SD-PET and vLD-PET
image was around 17 dB; the PSNR for the uLD-PET image was around 13 dB.
That is, the PSNR values for the set of uLD-PET images was around 0.66$\times$ that of the LD-PET images.  This variation in the PSNR values was
motivated by the work in~\cite{watson2005optimizing} that gives an example where the PET images' mean SNR reduced by a factor of around 0.66 when the
patients' body weight increased from around 60 kg to around 120 kg.

{ \color{black} \textbf{OOD data from different imaging protocols (OOD-Protocol).}  We use the dataset corresponding to the visual task experiments used for functional PET analysis in ~\cite{li2020analysis} and ~\cite{jamadar2019simultaneous}.  In brief, this dataset comprises T1w MRI, T2w MRI, and dynamic PET scans from six healthy subjects with mean age 24.3 years and standard deviation 3.8 years, including five females.  The scanner and MRI structural imaging protocols are the same as the data used for training all the DNN models (Section~\ref{sec:data}).  For PET, the scanning protocol involved bolus injection of 100 MBq of the radiotracer, which is significantly different from the training data (described in Section~\ref{sec:data}).  We consider the reconstructed PET images using the entire list-mode data as the reference PET image.  We generated a lower-quality PET image (input PET image) by using a part of the list-mode data such that the PSNR value, averaged across the entire dataset, between the reference PET image and input image was around 24 dB. }

{ \color{black} \textbf{OOD data from motion artifacts (OOD-Motion).}  Here, we use the dataset that is part of the study in \cite{chen2019mr}.  For OOD-Motion, we use the data corresponding to "Motion Controlled Study" from that study.  [18-F] FDG PET and structural MRI (T1w and T2w) data were acquired from a healthy volunteer.  For FDG-PET, a bolus of 110 MBq FDG was provided, and specific instructions pertaining to the head movement were provided at specific scan times.  We consider the reconstructed images using the (i)~entire list-mode data and (ii)~motion correction algorithm proposed in~\cite{chen2019mr} as the reference PET image.  We generated the lower-quality PET image (input PET image) by using part of the list-mode data.  Importantly, we did not perform any motion correction during or post-reconstruction.  The PSNR value averaged across the entire OOD-Motion dataset, between the reference PET image and the input PET image, was around 19 dB. }

{ \color{black} \textbf{OOD data from ADNI (OOD-ADNI: Alzheimer's Dementia; cross-scanner; multi-site; aged population data).}  We obtain a dataset from the Alzheimer's disease neuroimaging initiative (ADNI) database~\cite{weiner2017alzheimer}, which is a well-known publicly available dataset. We randomly selected data for 25 subjects (mean age 77 years and standard deviation 10.1 years, including 9 females) categorized as follows.  (i)~normal aging (2 subjects), (ii)~early mild cognitive impairment (EMCI, 4 patients), (iii)~mild cognitive impairment (MCI, 4 patients), (iv)~late mild cognitive impairment (LMCI, 8 patients), and (v)~dementia or AD (7 patients).  We obtained T1w, T2w, and [18-F] FDG PET images for all the subjects mentioned above.  The structural MRI images were acquired on a 1.5T and 3T scanners using 3D MPRAGE for T1w and FLAIR for T2w images with a resolution of 1mm$^3$ isotropic.  All the PET images were obtained at a resolution of 1.01 $\times$ 1.01 $\times$ 2.02 mm$^3$.  In comparison, the LD-PET and SD-PET data from OOD-Counts used for training the DNNs, were acquired on a 3T simultaneous PET-MRI scanner with a resolution of 1 mm$^3$ isotropic for MRI and 2.09 $\times$ 2.09 $\times$ 2.03 mm$^3$ for PET.  We registered and resampled all the PET and MRI images from the ADNI database to one of our training subjects to overcome differences in image resolution and image matrix dimensions.  For evaluation, we considered the provided reconstructed images as reference.  We retrospectively generated the degraded input images such that the PSNR value, averaged across the OOD-ADNI test set, between the reference PET image and the degraded input PET image was around 21 dB.  }

\subsection {Evaluation: Qualitative and Quantitative}

\begin{figure}[!t]
  \centerline{\includegraphics[width=1\columnwidth]{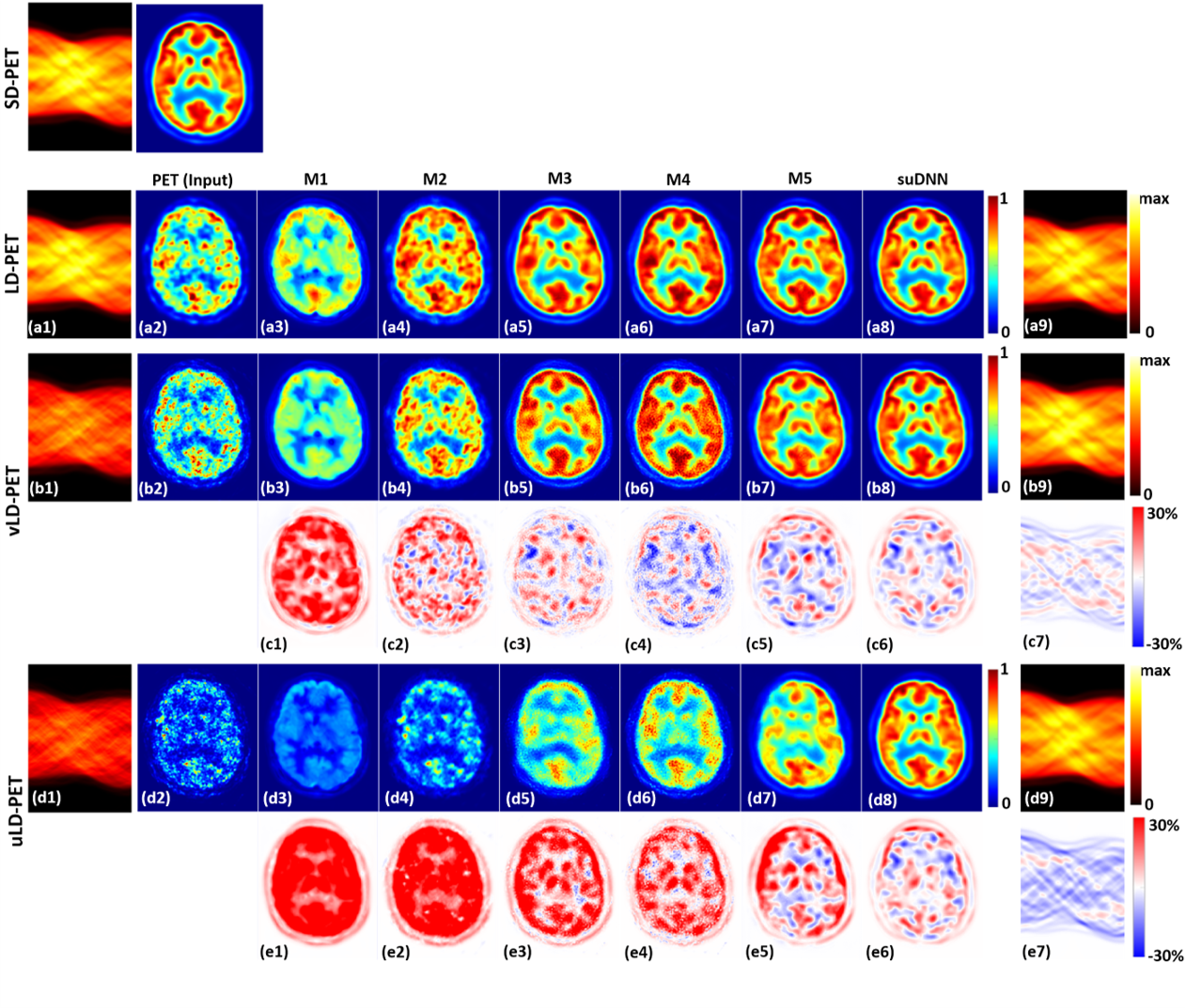}}
  \caption
      {
        {\bf Qualitative evaluation of the methods for three different levels of degradation of the input PET data: LD-PET (row a), vLD-PET (rows b
          and c), and uLD-PET (rows d and e). }
        {\color{black}
        The ground-truth SD-PET along with the corresponding sinogram are shown in the topmost row.
        }
        Panels (a1-a2) show the input LD-PET, (b1-b2) vLD-PET, and (d1-d2) uLD-PET sinograms and images; panels (a3-a8) the predicted images for all
        methods for LD-PET; panels (b3-b8) and (c1-c6) the predicted images and corresponding residual images (with respect to SD-PET) for vLD-PET;
        panels (d3-d8) and (e1-e6) the predicted images and corresponding residual images for uLD-PET as input; panels (a9-b9) and (d9) the sinograms
        of the predicted images (panels (a8, b8, and d8)); and panels (c7) and (e7) show the residuals of the predicted sinograms in comparison to the
        reference SD-PET sinogram.
      }
      \label{fig:compare_baselines_nls}
\end{figure}

Figure~\ref{fig:compare_baselines_nls} shows the predicted images from different methods across three different variations of the LD-PET data for a
representative subject.  The input PET images, i.e., LD-PET, vLD-PET, and uLD-PET, appear in Figures \ref{fig:compare_baselines_nls}(a2),
\ref{fig:compare_baselines_nls}(b2), and \ref{fig:compare_baselines_nls}(d2), respectively; the corresponding sinograms appear in Figures
\ref{fig:compare_baselines_nls}(a1), \ref{fig:compare_baselines_nls}(b1), and \ref{fig:compare_baselines_nls}(d1).
The DIP-based M1 (Figure~\ref{fig:compare_baselines_nls}(a3),(b3),(d3)) denoises the input LD-PET image.  However, as expected, being unsupervised and
with denoising as its focus, it is unable to enhance the low counts and performs poorly in predicting the FDG uptake in the reference SD-PET image.
Unlike M1, the ResNet-based M2 (Figure~\ref{fig:compare_baselines_nls}(a4),(b4),(d4)) is designed to predict the activity in the reference SD-PET
image.  However, even with the LD-PET input, it is unable to produce images with accurate textural features because of several possible factors. One
factor is that M2's design cannot leverage the information in the MRI image.
M2 relies on a standard ResNet architecture that employs short-range skip connections compared to longer-range hierarchically-designed skip
connections in suDNN's U-Net architecture.
Methods M3 (Figure~\ref{fig:compare_baselines_nls}(a5),(b5),(d5)) and M4 (Figure~\ref{fig:compare_baselines_nls}(a6),(b6),(d6)), which rely on
predicting the residual images as output, produce realistic SD-PET images when using LD-PET as the input.  However, when using vLD-PET and uLD-PET as
inputs, both M3 and M4 show some residual noise in the images despite reasonably recovering the contrast and texture similar to the SD-PET image.
M4 improves over the loss in contrast shown by M3, emphasizing the contribution of the multimodal MRI input.
M5, which is GAN-based, shows superior performance with LD-PET (Figure~\ref{fig:compare_baselines_nls}(a7)), showing little degradation (in terms of
contrast and certain structures like the sulci and gyri) with vLD-PET (Figure~\ref{fig:compare_baselines_nls}(b7)), and does {\em not} predict the
desired texture and contrast when using uLD-PET as input (Figure~\ref{fig:compare_baselines_nls}(d7)).
On the other hand, our suDNN shows superior prediction across varying input quality (Figure~\ref{fig:compare_baselines_nls}(a8),(b8),(d8)). Compared
to other baselines, suDNN's results show more realistic texture and contrast, and reduced magnitudes in the differences between the predicted and the
reference SD-PET images (Figure~\ref{fig:compare_baselines_nls}(c6),(e6)).
For our suDNN, the sinograms of the predicted images (Figures~\ref{fig:compare_baselines_nls}(a9),(b9),(d9)) demonstrate little difference across OOD
variations in input image quality, which is in agreement with the quality of the predicted images obtained with different low-dose inputs.  The
residual images between the sinograms of the predicted images and that of the reference image SD-PET corresponding to the inputs vLD-PET and uLD-PET
are shown in Figure~\ref{fig:compare_baselines_nls}(c7)-(d7).
\begin{figure}[!t]
  \centerline{\includegraphics[width=0.8\columnwidth]{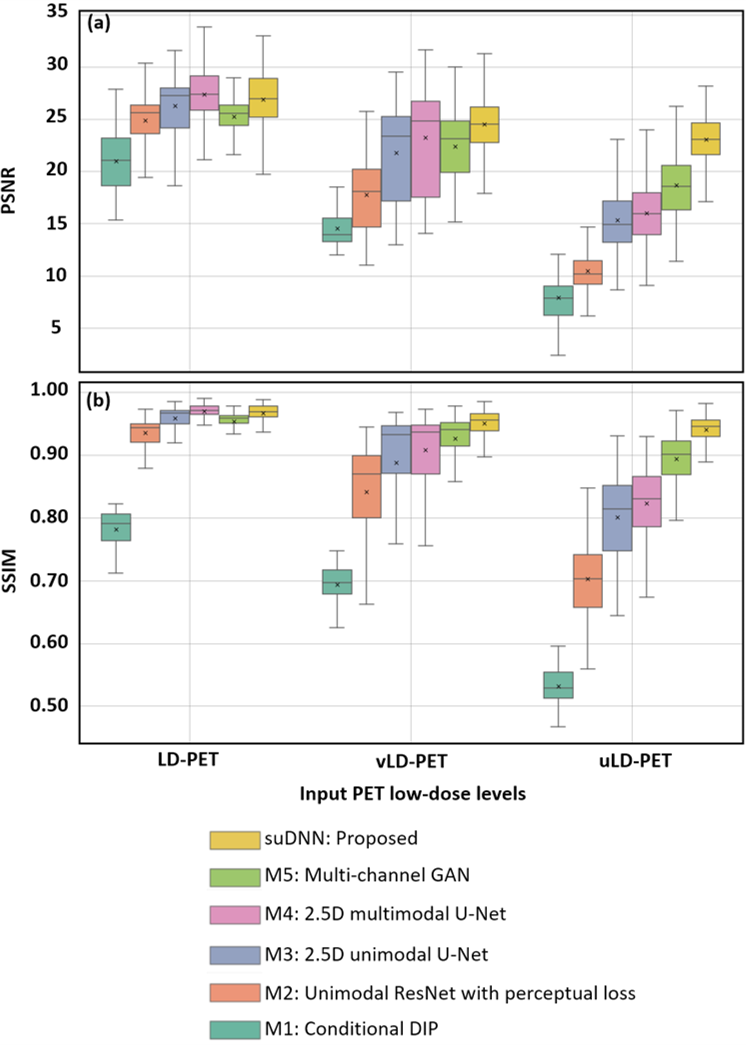}}
  \caption
      {
      {\bf Quantitative evaluation of the methods for three different levels of degradation of the input PET data: LD-PET, vLD-PET, and uLD-PET.}
      {\bf (a)} PSNR and {\bf (b)} SSIM values for the predicted PET images on 100 brain slices for each test set.
      \revise
      {
      The plots depict performance on the test-set averaged over a 3-fold cross-validation scheme. 
      }
    }
    \label{fig:quant_baselines}
\end{figure}

\begin{figure}[!t]
  \centerline{\includegraphics[width=1\columnwidth]{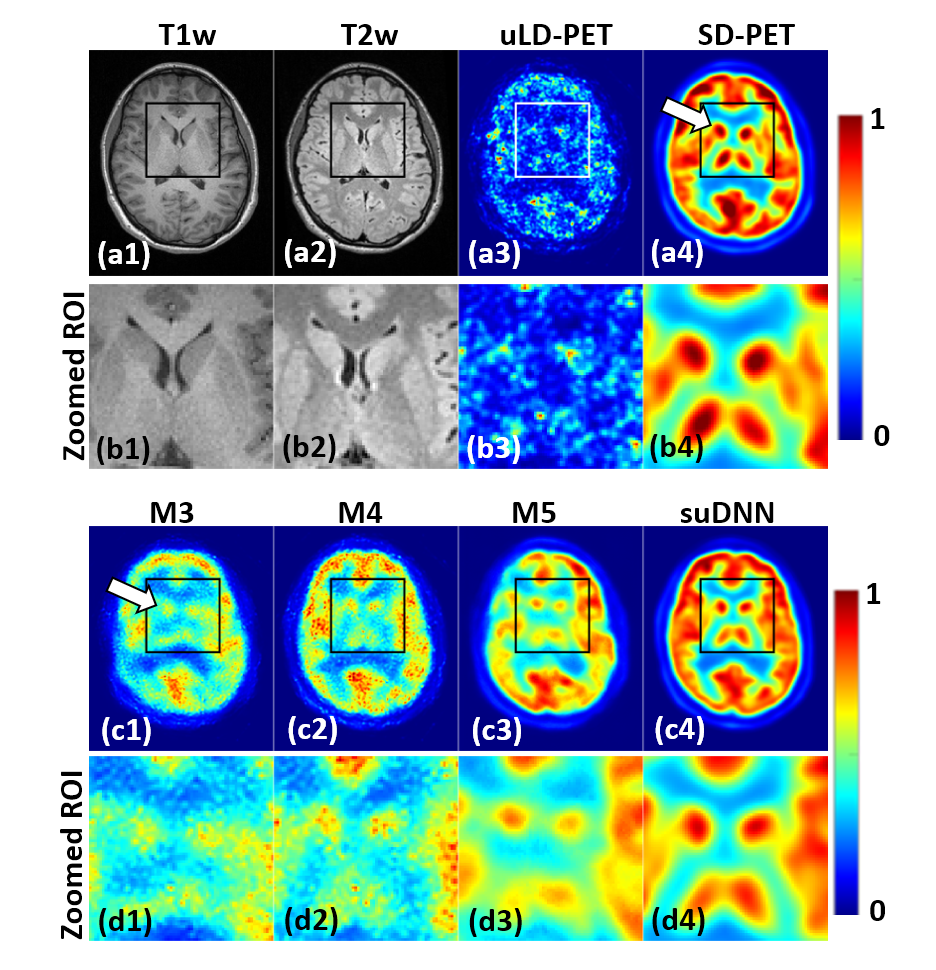}}
  \caption
      {
        {\bf Zoomed ROIs of the input, reference, and predicted images with for the case of uLD-PET.}
        {\bf (a1)--(a3)}: input T1w MRI, T2w MRI, and uLD-PET image images.
        {\bf (a4)}: reference SD-PET image.
        {\bf (c1)--(c4)}: predicted images from the methods M3--M5 and the proposed suDNN method.
        {\bf (d1)--(d4)} corresponding zoomed regions.
      }
      \label{fig:zoomed}
\end{figure}

Figures~\ref{fig:quant_baselines}(a)-(b) show quantitative plots with PSNR and SSIM values 
\ud{
  averaged over the 100 slices of every subject from the test set in 3-fold cross validation (18 patients for training, 4 for validation, 6 for testing)
}
for different kinds of PET image inputs, i.e., LD-PET, vLD-PET, and uLD-PET.
As the input quality degrades, all methods show a drop in performance.  Nevertheless, our method shows the most graceful degradation (around 3.5 dB
with uLD-PET).  On the other hand, the other methods show a severe loss in their performance with uLD-PET, e.g., around 7~dB for M5, around 10~dB for
M4, and around 11~dB for M3.
A similar trend can be observed in the SSIM plot (Figures~\ref{fig:quant_baselines}(b)).  While our method shows a degradation of around $0.02$ with
uLD-PET as input as compared to LD-PET as the input, other methods show a severe decrease in SSIM values with uLD-PET, e.g., around 0.04 for M5,
around 0.13 for M4, and around 0.1 for M3.  Thus, with LD-PET as input, the performance of suDNN is comparable to M3--M5; nevertheless, as the input
quality degrades, suDNN significantly outperforms all other methods demonstrating substantially higher robustness/insensitivity to OOD data.
We conducted paired {\em t}-test for SSIM and PSNR values for all methods for the three low-dose inputs. The improvement using our suDNN method was
found to be statistically significant ($p$-value $\ll 0.001$) in comparison to all other methods (M1--M5) at all input quality levels (LD-PET,
vLD-PET, and uLD-PET).

For the results corresponding to uLD-PET input in Figure~\ref{fig:compare_baselines_nls}, we carefully analyze the predicted images along with the
input and the reference PET images. The zoomed region of interest (ROI) includes the caudate, putamen, and thalamus.  The caudate nucleus shows
hyperintensity in the SD-PET image (highlighted using the white arrow in Figure~\ref{fig:zoomed}(a4)) that is {\em not} the case in the uLD-PET image
(Figure~\ref{fig:zoomed}(a3)-(b3)).  The unimodal DNN M3 (Figure~\ref{fig:zoomed}(c1)-(d1)) severely underestimates the uptake in the caudate and the
thalamus regions. Although our suDNN (Figure~\ref{fig:zoomed}(c4)-(d4)) provides the best estimate of the predicted images, other multimodal DNN
methods like M4 and M5 (Figure~\ref{fig:zoomed}(c2)-(d2) and Figure~\ref{fig:zoomed}(c3)-(d3)) do show some recovery of the hyperintensity in the
caudate and thalamus regions compared to M3.  This demonstrates the importance of including the MRI structural image in the input, where the results
(Figure~\ref{fig:zoomed}(a1)-(b1) and Figure~\ref{fig:zoomed}(a2)-(b2)) distinctly show the subcortical nuclei in the cerebrum.
\begin{figure}[!t]
  \centering
  \includegraphics[width=0.93\columnwidth]{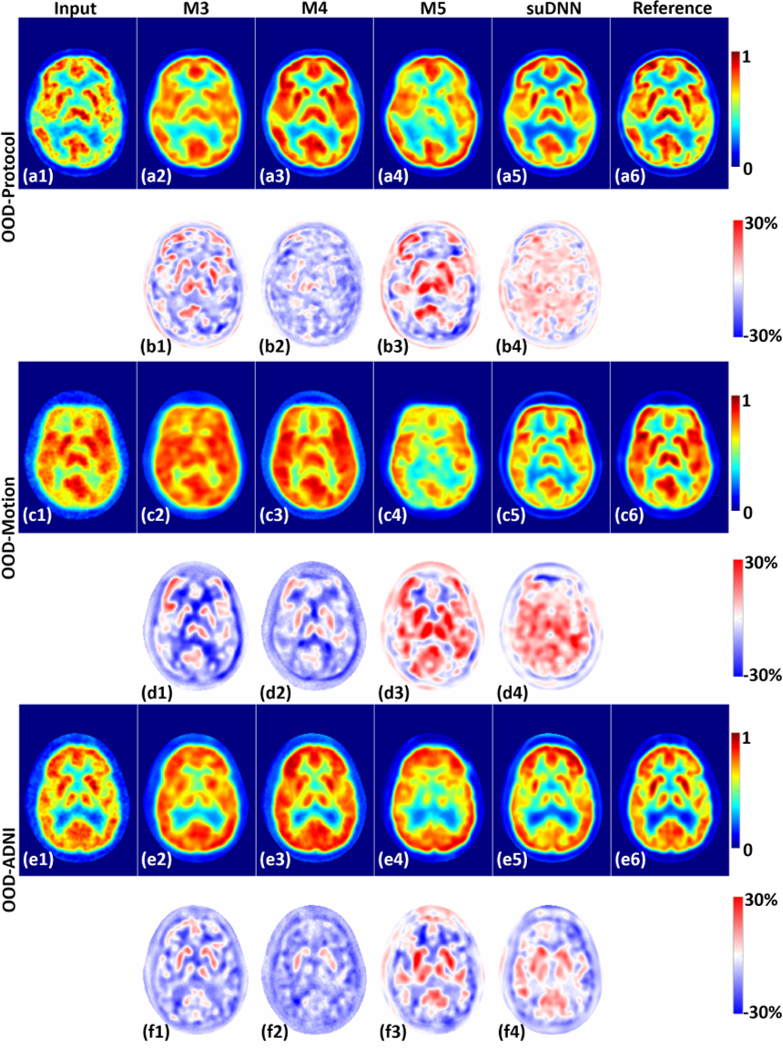}
  \caption
      {
      \revise
      {
        {\bf Qualitative evaluation of the methods for three additional types of OOD data: OOD-Protocol (rows a and b), 
        OOD-Motion (rows c and d), 
        and OOD-ADNI (rows e and f). }
        Panels (a1, c1, and e1) correspond to the input PET images, 
        and (a6, c6, and e6) correspond to the reference PET images.
        Columns 2--5 show the predicted images and residuals for the methods: M3 (column 2), M4 (column 3), M5 (column 4), and suDNN (column 5).
        }
        }
    \label{fig:qualitative_other_ood}
\end{figure}
\begin{figure}[!t]
  \centerline{\includegraphics[width=\columnwidth]{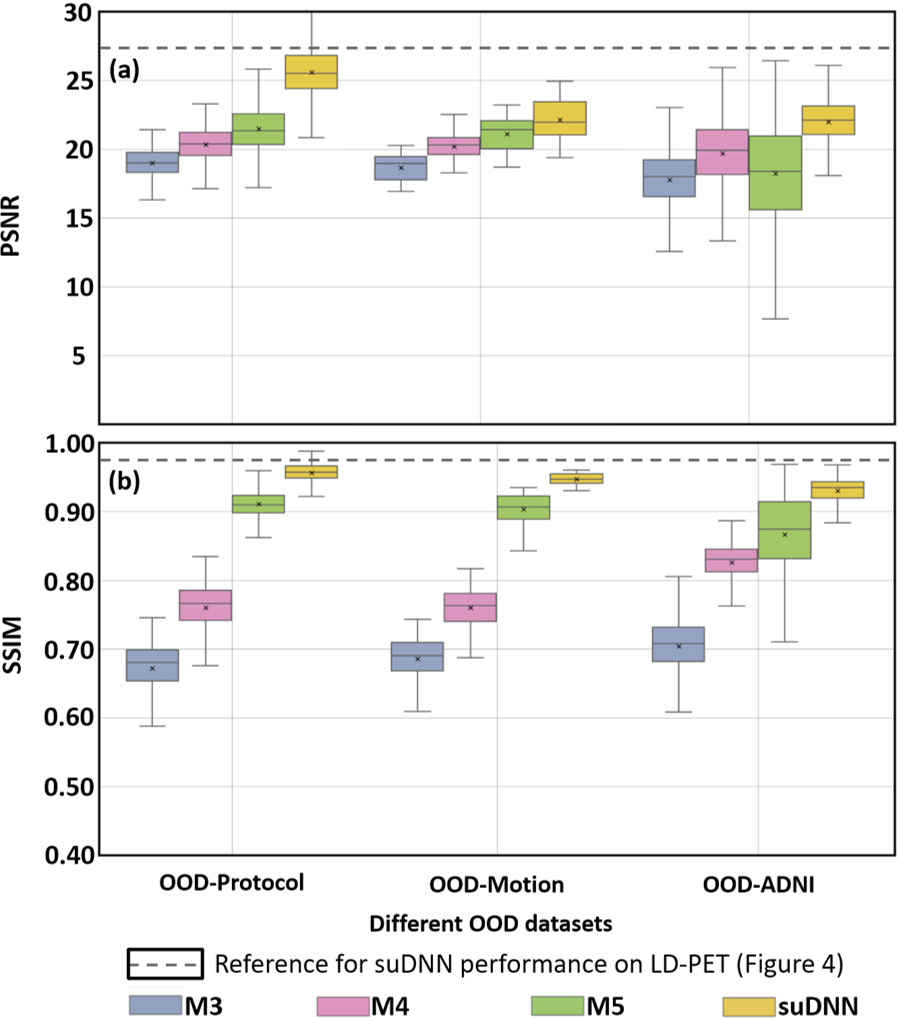}}
  \caption
      {
      \revise{
      {\bf Quantitative evaluation of the methods for three additional OOD data:
      OOD-Protocol (6 subjects), OOD-Motion (1 subject), and OOD-ADNI (25 subjects).}
      {\bf (a)} PSNR and {\bf (b)} SSIM values for the predicted PET images on 100 brain slices
      for each test subject under each case of OOD.
      The dotted lines represent the median PSNR (in (a)) and SSIM (in (b)) values  
      obtained from the performance of suDNN on LD-PET data (part of OOD-Counts) in Figure\ref{fig:quant_baselines}.
    }
    }
    \label{fig:quant_other_ood}
\end{figure}

{
\color{black}
Figure~\ref{fig:qualitative_other_ood} shows the predicted PET images and residuals
for the models M3, M4, M5, and suDNN on
three additional OOD datasets OOD-Protocol, OOD-Motion, and OOD-ADNI.
For OOD-Protocol, while M3 (Figure~\ref{fig:qualitative_other_ood} (a2))
shows under-estimation (compared to the reference) in the entire brain region, 
M4 (Figure~\ref{fig:qualitative_other_ood} (a3)) shows increased activity across the entire brain.
On the other hand, suDNN (Figure~\ref{fig:qualitative_other_ood} (a5)) closely matches the activity distribution across brain regions without severe under- or over- estimation,  
yielding the least residual magnitudes (Figure~\ref{fig:qualitative_other_ood} (b4)).
For OOD-Motion, unlike OOD-Protocol, 
both M3 and M4 (Figure~\ref{fig:qualitative_other_ood} (c2) and (c3))
show increased activation across the entire brain region and are also unable to recover certain anatomical structures (e.g., caudate nuclei).
Relatively, suDNN (Figure~\ref{fig:qualitative_other_ood} (c5)) is able to closely match the activity distribution across brain regions.
For OOD-ADNI too,  
suDNN (Figure~\ref{fig:qualitative_other_ood} (e5)) provides substantially
improved images \revisesec{compared to other methods} (Figure~\ref{fig:qualitative_other_ood} (e2)--(e4)) 
with the least residual magnitudes. 
Across all the three OOD datasets, M5 (Figure~\ref{fig:qualitative_other_ood} (a4, c4, and e4) and (b3, c3, and f3))
is unable to recover certain subcortical structures.
Nevertheless, unlike M3 and M4, it does {\em not} suffer from severe under- or over-estimation. 
Thus, across the three additional OOD datasets discussed here, our proposed method (suDNN), shows reliable
(i)~activity estimation and (ii)~anatomical structure restoration
compared to M3, M4, and M5.

%
Figures~\ref{fig:quant_other_ood}(a)-(b) show quantitative plots with PSNR and SSIM values 
for 100 slices of every subject for each of the three 
additional OOD datasets: OOD-Protocol, OOD-Motion, and OOD-ADNI.
The dotted lines in both the plots indicate the median PSNR and SSIM values 
of suDNN evaluated on LD-PET dataset (part of OOD-Counts) from Figure~\ref{fig:quant_baselines}.
Across all the three OOD datasets, our method performs 
significantly better (around 4 dB for OOD-Protocol,
and around 1.5 dB for OOD-Motion and OOD-ADNI) than M3, M4, and M5.
On OOD-Protocol, our method's performance is comparable to its corresponding performance on the LD-PET test data (in OOD-Counts).
A similar trend can be observed in the SSIM plot (Figures~\ref{fig:quant_other_ood}(b)).  
While our method on OOD-Protocol shows comparable SSIM values compared to LD-PET data in OOD-Counts,
it shows a slight degradation of around 0.1 and 0.2
for OOD-Motion and OOD-ADNI, respectively.
}

\subsection {Ablation Studies: Qualitative and Quantitative}

We perform an ablation study to analyze the contribution from different components in the proposed DNN.
To this end, consistent with the prior works in this domain, we found that using a 2.5D-input based training scheme provided substantially improved
results in comparison to using 2D-only training. Moreover, as evident from the results in Figures~\ref{fig:compare_baselines_nls}--\ref{fig:zoomed},
M3 and M4 that rely on predicting the residual between the LD-PET and the SD-PET images, are {\em not} robust to OOD acquisitions.
Hence, to evaluate the importance of multiple components in the proposed suDNN framework, we evaluate {\em four} other ablated versions of suDNN,
i.e., suDNN-Ablated1, suDNN-Ablated2, suDNN-Ablated3, and suDNN-Ablated4.

\begin{itemize}
\item %
  {\bf suDNN-Ablated1: 2.5D unimodal U-Net.} We define a basic DNN that includes a U-Net architecture (similar to~\cite{subtle}) with a unimodal
  input, but with a modified output such that it directly maps to the PET image instead of estimating the residual between the input LD-PET and the
  reference SD-PET image (as in M3).  suDNN-Ablated1 is trained using the 2.5D scheme, penalizing the mean-squared error in the image space, say
  $\mathcal{L}_\text{I} (\widehat{Y}, U^\text{SD})$, between the predicted and the reference images.
\item %
  {\bf suDNN-Ablated2: 2.5D multimodal U-Net.} We modify the DNN suDNN-Ablated1 by replacing the unimodal input with a multimodal input including
  multi-contrast MRI images, retaining the same loss function as suDNN-Ablated1.
\item %
  {\bf suDNN-Ablated3: 2.5D multimodal U-Net with manifold loss.} In addition to the loss $\mathcal{L}_\text{I}$, this DNN includes a learned
  manifold-based loss $\mathcal{L}_\text{E} (\widehat{Y}, U^\text{SD})$ similar to the perceptual loss in~\mycite{johnson2016perceptual} or the
  manifold-based loss in~\mycite{upadhyay2019robust}; thus, the total loss is $\mathcal{L}_\text{I} + \lambda_\text{E} \mathcal{L}_\text{E}$, where
  the free parameter $\lambda_\text{E} \in \mathbb{R}^{+}$ controls the weight of the loss term $\mathcal{L}_\text{E}$.  The learned-manifold based
  loss relies on learning an autoencoder trained using the set of SD-PET images.  The loss function $\mathcal{L}_\text{E}$ penalizes the differences
  between the encodings obtained by applying the encoder $\Phi_\text{E}$ (from learned autoencoder) to the predicted PET and reference SD-PET
  images. That is, $\mathcal{L}_\text{E} (\widehat{Y}, U^\text{SD}; \Phi_\text{E}) := \| \Phi_\text{E} (\widehat{Y}) - \Phi_\text{E} (U^\text{SD})
  \|_F^2$, where $\|\cdot\|_F$ represents the Frobenius tensor norm.
\item  
  {\bf suDNN-Ablated4: 2.5D multimodal U-Net with physics-based loss.}  Instead of the learned-manifold loss in suDNN-Ablated3 , suDNN-Ablated4 uses a
  sinogram-space loss $\mathcal{L}_\text{S}$ given as $\mathcal{L}_\text{S} := \| \mathcal{S} \widehat{Y} - \mathcal{S} \widehat{U}^\text{SD} \|_F^2$.
  Thus, the total loss for suDNN-Ablated4 is $\mathcal{L}_\text{I} + \lambda_\text{S} \mathcal{L}_\text{S}$, where $\lambda_\text{S} \in
  \mathbb{R}^{+}$ controls the strength of $\mathcal{L}_\text{S}$.
\end{itemize}
The free parameters $\lambda_\text{E}$ and $\lambda_\text{S}$ are automatically tuned using the validation set; in this paper, they take the values
$\lambda_\text{E} = 0.002$ and $\lambda_\text{S} = 0.003$.

\begin{figure}[!t]
  \centerline{\includegraphics[width=0.9\columnwidth]{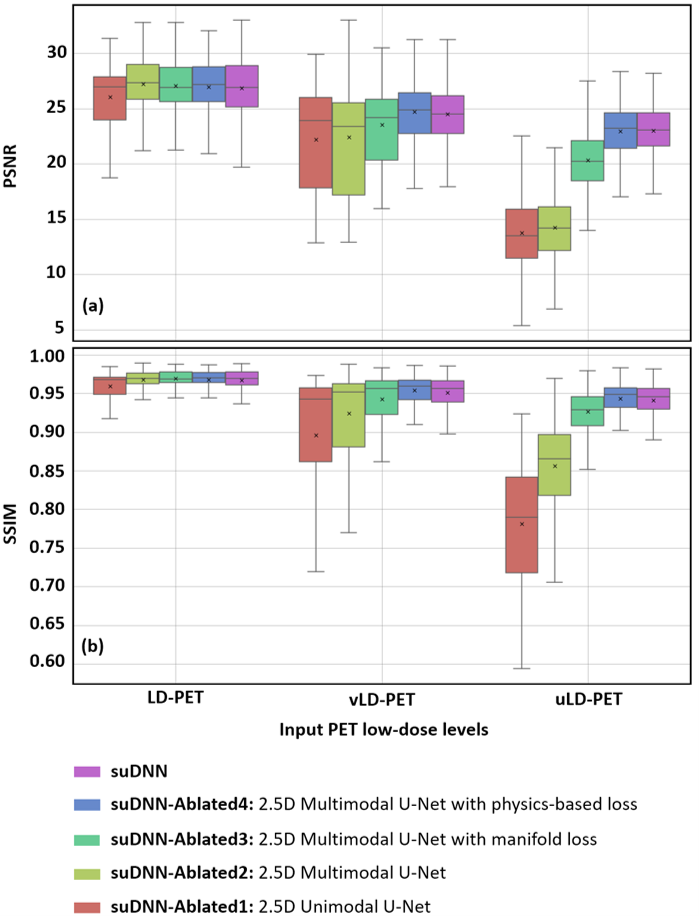}}
  \caption
      {
        {\bf Ablation Study: Quantitative evaluation for the ablation study at three different levels of degradation of the input PET data: LD-PET,
          vLD-PET, and uLD-PET.}
        {\bf (a)} PSNR and {\bf (b)} SSIM values for predicted SD-PET images, on 100 brain slices in every test set.
      }
      \label{fig:ablation_study}
\end{figure}

Figure~\ref{fig:ablation_study} shows quantitative evaluation of the DNNs in the ablation study for the input PET images LD-PET, vLD-PET, and uLD-PET.
Similar to the results in Figure~\ref{fig:quant_baselines}, DNNs with a multimodal input improve substantially over DNNs with unimodal input (suDNN,
suDNN-Ablated2, suDNN-Ablated3, and suDNN-Ablated4 better than suDNN-Ablated1).
Inclusion of the learned manifold-based loss $\mathcal{L}_\text{E} (\cdot)$, in addition to the image space loss $\mathcal{L}_\text{I} (\cdot)$, for
suDNN-Ablated3 provides improved robustness over suDNN-Ablated2 and suDNN-Ablated1.
Further, suDNN-Ablated4 that includes a physics-based loss instead of the learned manifold-based loss in suDNN-Ablated3 shows significant improvement
over suDNN-Ablated3 with vLD-PET and uLD-PET.
Finally, the proposed suDNN that models uncertainty in both image and sinogram space, provides comparable performance to suDNN-Ablated4, but
significantly better than suDNN-Ablated1, suDNN-Ablated2, and suDNN-Ablated3 at higher levels of degradation of the input.
In addition to providing improved accuracy and robustness to OOD data over other methods, the predicted variance image $\widehat{C}$ from the proposed
DNN can potentially be useful for quantifying the uncertainty in the predicted images discussed in Section~\ref{sec:uncertainty}.

\begin{figure}[!t]
  \centerline{\includegraphics[width=1\columnwidth]{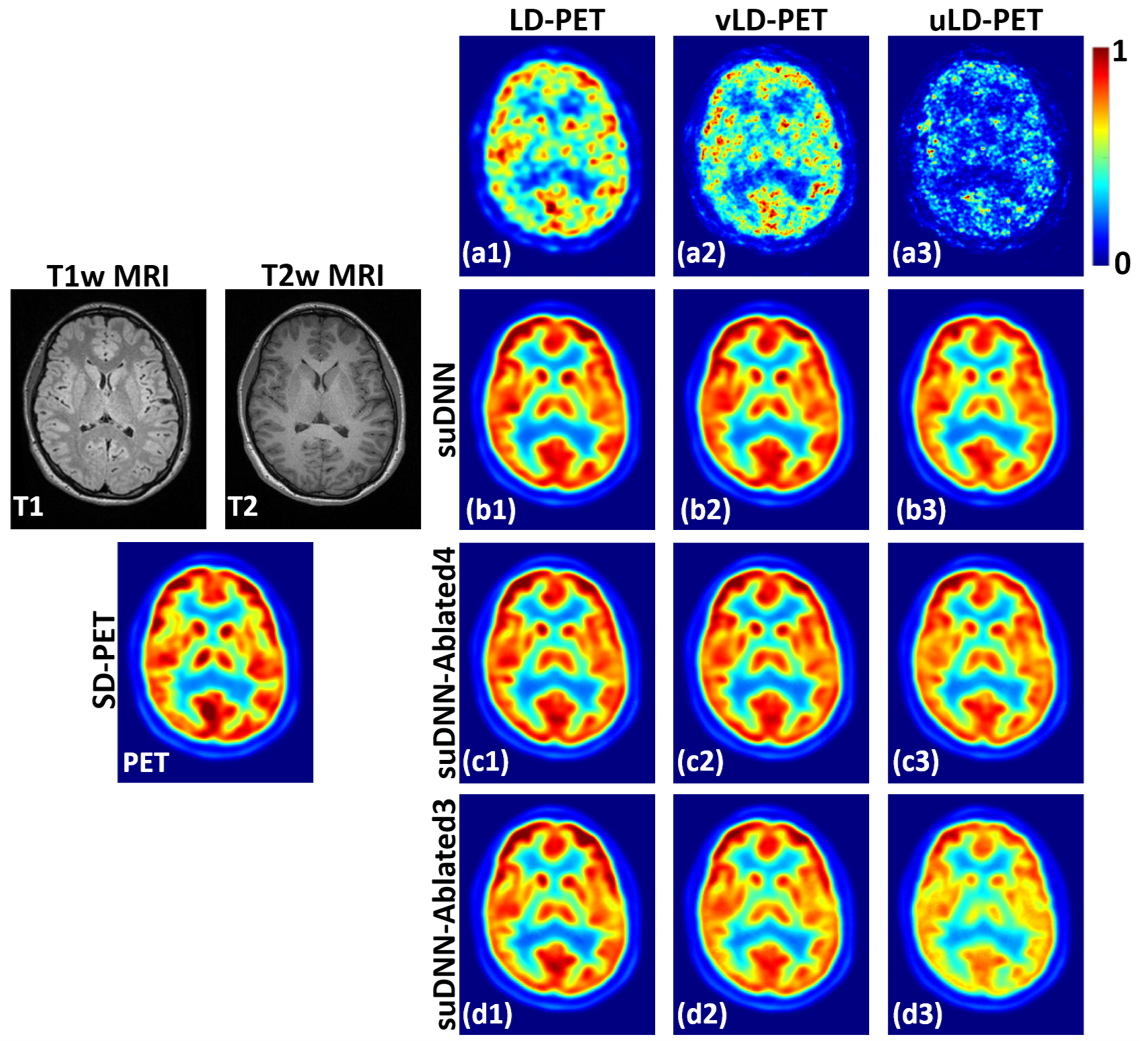}}
  \caption
      {
        {\bf Results of DNNs in the ablation study with input PET images: LD-PET, vLD-PET, uLD-PET.}
        Variations in input PET {\bf (a1)--(a4)}: LD-PET, vLD-PET, uLD-PET, respectively.
        Predicted images using varying levels of PET input from:
        {\bf (b1)--(b3)}: suDNN,
        {\bf (c1)--(c3)}: suDNN-Ablated4, and
        {\bf (d1)--(d3)}: suDNN-Ablated3.
      }
      \label{fig:visualize_ablation}
\end{figure}

\begin{figure}[!t]
  \centerline{\includegraphics[width=1\columnwidth]{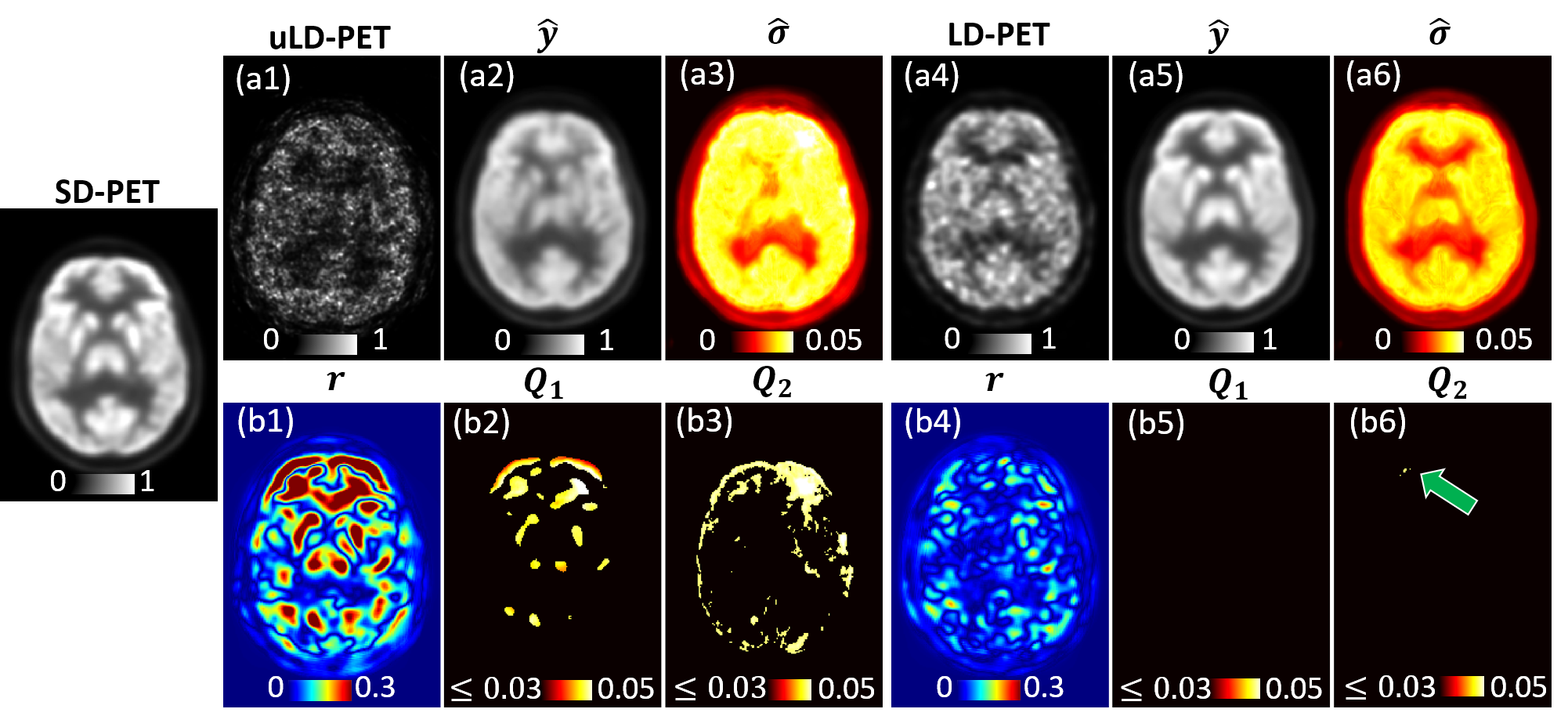}}
  \caption
      {
        {\bf Utility of Uncertainty Maps; Columns 1--3: uLD-PET as input; Columns 4--6: LD-PET as input.}
        {\bf (a1) and (a4)}: Input images uLD-PET and LD-PET.
        {\bf (a2) and (a5)}: Predicted PET images $\widehat{Y}$.
        {\bf (a3) and (a6)}: Predicted per-voxel standard deviation image $\sqrt{\widehat{C}}$.
        {\bf (b1) and (b4)}: Image $r$ showing magnitudes of per-voxel residuals in 
        $U^\text{SD} - \widehat{Y}$.
        {\bf (b2), (b5)}: Quantification map $Q_1(\widehat{\sigma}; r, \delta_\text{R})$;
        {\bf (b3), (b6)}: Quantification map $Q_2(\widehat{\sigma}; \delta_\text{U})$.
        \revise{
          We observe that the regions with large values in $Q_1$ are subsumed within region with large values in $Q_2$.
        }
      }
      \label{fig:uncertainty}
\end{figure}

Figure~\ref{fig:visualize_ablation} provides visual comparison of the output SD-PET images from the ablated suDNN versions suDNN-Ablated3,
suDNN-Ablated4, and the proposed suDNN, for the input PET images (i)~LD-PET, (ii)~vLD-PET, and (iii)~uLD-PET.  For the LD-PET and vLD-PET inputs, the
predicted PET images from suDNN-Ablated3 (Figure~\ref{fig:visualize_ablation}(d1)-(d2)) are closer to that of suDNN-Ablated4 and suDNN
(Figure~\ref{fig:visualize_ablation}(c1),(b1) and Figure~\ref{fig:visualize_ablation}(c2),(b2)).  However, suDNN-Ablated3 shows substantial
degradation with uLD-PET as input (Figure~\ref{fig:visualize_ablation}(d3)).  The outputs of suDNN-Ablated4
(Figure~\ref{fig:visualize_ablation}(c1)--(c3)) are very similar to that of suDNN (Figure~\ref{fig:visualize_ablation}(b1)--(b3)).  This 
\revise
{
emphasises
}
that modeling uncertainty in both the image space and the sinogram space, need {\em not} hamper the image quality.

\subsection{Utility of Uncertainty Maps}
\label{sec:uncertainty}

We now analyze the uncertainty maps produced by the proposed suDNN with the inputs uLD-PET and LD-PET, and how to extract useful information from the
same.
For the input PET images uLD-PET and LD-PET (Figure~\ref{fig:uncertainty}(a1) and Figure~\ref{fig:uncertainty}(a4)), the network produces the
predicted images (Figure~\ref{fig:uncertainty}(a2) and Figure~\ref{fig:uncertainty}(a5), respectively), along with the per-voxel variances
$\widehat{C}$. For improved visualization, we show the uncertainty maps, i.e., per-voxel square-root of the variance maps, $\widehat{\sigma} :=
\sqrt{\widehat{C}}$ (Figure~\ref{fig:uncertainty}(a3) and Figure~\ref{fig:uncertainty}(a6)).
We define two global thresholds to identify pixels with high uncertainty and high residual magnitudes, i.e., threshold $\delta_\text{U}$ for the
predicted uncertainty image and threshold $\delta_\text{R}$ for the residual-magnitude image.  That is, voxel locations with residual-magnitude values
$r \geq \delta_\text{R}$ indicate sub-optimal reconstruction, and voxel locations with $\widehat{\sigma} \geq \delta_\text{U}$ indicate predictions
with high uncertainty.  Subsequently, we threshold the residual-magnitude image $r$ and the uncertainty image $\widehat{\sigma}$ to get two binary
masks, namely, BM1 and BM2. We tune the values for the global thresholds empirically to $\delta_\text{R} = 0.25$ and $\delta_\text{U} = 0.03$,
respectively.
Finally, to improve the utility of the uncertainty maps, we generate two quantification maps:
(i)~$Q_1 (\widehat{\sigma}; r, \delta_\text{R})$ (Figure~\ref{fig:uncertainty}(b2) and Figure~\ref{fig:uncertainty}(b5)), obtained by applying the
binary mask BM1 on $\widehat{\sigma}$, and
(ii)~$Q_2 (\widehat{\sigma}; \delta_\text{U})$ (Figure~\ref{fig:uncertainty}(b2) and Figure~\ref{fig:uncertainty}(b5)), obtained by applying the
binary mask BM2 on $\widehat{\sigma}$.
As expected, the map $Q_1$ with the LD-PET input has substantially fewer non-zero values, compared to the map $Q_1$ obtained with uLD-PET as input.  A
similar trend is observed for the map $Q_2$.  Thus, as expected, suDNN's prediction from uLD-PET as input shows higher uncertainty compared to its
prediction from LD-PET as input.  Notably, the high-intensity values in the map $Q1$ agree with the high-intensity values in the map $Q2$; this
implies that regions with high residual magnitudes correspond to regions with high uncertainty in the predicted images.
In this way, the map $Q_2$ (available at inference) might act as a proxy for the prediction error (i.e., residual-magnitude map $Q_1$ that is
unavailable at test time) while inferring a PET reconstruction from test data.

\section{Discussion and Conclusion}
\label{sec:discussion}

This paper presents a novel sinogram-based and uncertainty-aware DNN framework, namely, suDNN, for estimating SD-PET images from LD-PET images, and
given the associated multi-contrast MRI, in simultaneous PET-MRI systems.
Specifically, we learn the mapping using LD-PET images associated with a DRF of 180~$\times$, and show that the learned mapping is robust to practical
OOD degradations in the data, i.e., PET data with further reduction in counts leading to vLD-PET \revise{(10$\times$)} and uLD-PET \revise{(100$\times$)} images, which realistically model the SNR
variation of the OSEM-reconstructed PET images in practical scenarios~\cite{watson2005optimizing}.
{
\color{black}
Furthermore, given the trained model on LD-PET images, 
we evaluated the performance on three additional OOD datasets
capturing variation in data due to several factors such as:
FDG infusion protocol and dose (OOD-Protocol), 
subject motion (OOD-Motion), and
age, pathology, multi-site, cross-scanner data acquisition (OOD-ADNI).
}
Compared to several existing methods, empirical evidence shows suDNN to be more robust 
(Figures~\ref{fig:compare_baselines_nls}, ~\ref{fig:quant_baselines},
{
\color{black}
~\ref{fig:qualitative_other_ood}, and \ref{fig:quant_other_ood})  
}.
Furthermore, unlike other methods, suDNN models the per-voxel heteroscedasticity during learning and inference
and, thereby provides useful information about the uncertainty in the predicted images.
Improving the robustness of the learned DNN to
effectively handle a wide spectrum of OOD variations reduces the number of learned DNN models required for deployment (for a particular combination of
a tracer and an anatomical region).

This is the first work, to the best of our knowledge, to include a PET-physics based (sinogram domain) loss function for enhancing LD-PET images. The
ablation study (Figure~\ref{fig:ablation_study}) shows that inclusion of the physics-based transform-domain loss function improves the robustness to
OOD data in the form of lower counts. This finding is consistent with findings in undersampled MRI reconstruction that show that modeling penalties in
the transform/k-space domain improve the performance of the DNN~\mycite{DAGAN}.
This is also the first work towards the modeling and quantification of the uncertainty in the predicted SD-PET images from LD-PET images.

Evaluating the performance \revise{of} suDNN as well as other state-of-the-art DNNs showed that the unimodal (M3) and the multimodal (M4) residual-predicting
U-Net DNNs are far less robust to OOD input data in the form of vLD-PET and uLD-PET.
Although, with vLD-PET as input, the multimodal GAN-based M5 improves over M3 and M4, it underestimates the SD-PET contrast with uLD-PET as input.
Unlike the empirical analysis in previous works that employ test data and training data having well-matched distributions, we evaluate the robustness
of all trained DNNs to OOD PET acquisitions leading to lower photon counts (at test time). While we train the DNN using LD-PET images and evaluate the
learned model on vLD-PET and uLD-PET images, one could also perform similar studies by learning the DNN model at some other specific level of image
quality and evaluating the learned model at the remaining levels.

\begin{figure}[!t]
  \centerline{\includegraphics[width=\columnwidth]{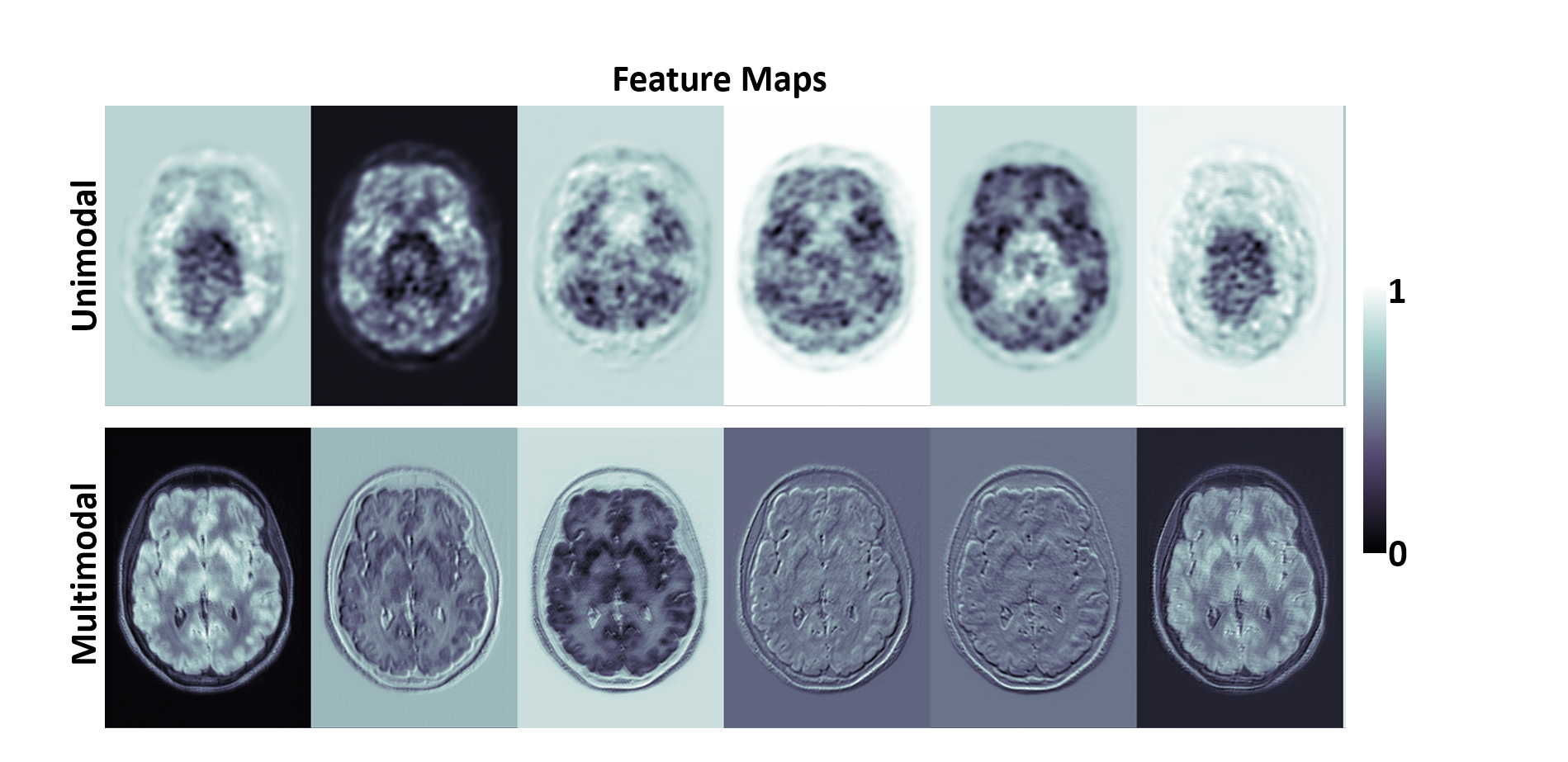}}
  \caption
      {
        {\bf Feature maps obtained from initial layers of the proposed network with unimodal (PET only) and multimodal inputs (PET and multi-contrast
          MRI both)}.
        Example 6 feature maps (out of 64) selected at the output of the second layer of the DNN are shown.  Feature maps have been normalized for
        better comparison between the unimodal and the multimodal case.
      }
      \label{fig:filtered_images}
\end{figure}

The use of multi-contrast MRI images as multi-channel input (in M4, M5, and suDNN) provides a substantial improvement over unimodal PET-only inputs
(M2 and M3), which is consistent with the findings of the other works for this problem~\mycite{wang20183d,subtleRadiology}.

{
\color{black}
The evaluation study on the three additional OOD datasets showed that even without additional training with the new data, our model is able to better adapt to restore structures and 
activity distribution both.
For the OOD data arising from the same imaging center,
with variations in PET (e.g., due to counts, population, subject motion, reconstruction pipeline),
but retaining the same MRI contrasts (T1w MRI and T2w MRI),
the predicted PET images closely matched the reference PET images.
However, for better generalizability spanning across scanners, 
imaging protocol, differences in PET radiotracer infusion protocol, 
as in the case of OOD-ADNI dataset, 
performance of all DNN models can be improved by further training of the pre-trained DNNs
using a few samples from the newer imaging sites. 
}

Furthermore, the ablation studies (Figure~\ref{fig:ablation_study} and Figure~\ref{fig:visualize_ablation}) show that DNNs that include multimodal
inputs as well as transform-domain losses (e.g., manifold loss or sinogram loss) produce better outputs even with reduced counts in the PET images.
The results in Figure~\ref{fig:visualize_ablation} also emphasize the importance of the information from the PET images for improved accuracy.
For DNN models that employ multimodal input (in suDNN and other works), e.g., including multi-contrast MRI as input, the non-PET modalities help
improve the prediction by infusing reliable information in the form of inter-modality statistical dependencies. In this context, to retain the
interpretation of PET imaging as quantitative imaging, a recalibration mechanism based on relative reduction in activity~\mycite{ouyang2019ultra} may
be needed, which is a part of our future work.

To analyze the contribution of the multimodal inputs in comparison to the unimodal (PET-only) inputs, we visualize the feature maps obtained from an
initial layer (second layer) of the DNN, trained with unimodal and with multimodal inputs, while maintaining the same network architecture.
Figure~\ref{fig:filtered_images} shows that the feature maps obtained using the multimodal inputs show anatomical features more clearly, compared to
the unimodal case, as expected.
We demonstrated the potential utility of the generated uncertainty maps (Figure~\ref{fig:uncertainty}) by defining global thresholds in terms of
residual magnitude and uncertainty values obtained in the experiments. Future work calls for defining these thresholds in terms of physically
meaningful values.
\revisesec{
There could be other approaches such as in \cite{dropout} 
that model dropout within a variational-learning framework for uncertainty estimation,
which may result in non-trivial extensions and modifications of the proposed suDNN framework.
However, studying such approaches is beyond the scope of this work.}

Some aspects of the analysis within this paper can improve in future works.
First, suDNN uses a 2.5D-style input instead of full 3D volumes.  In the future, we plan to accommodate training using 3D images, which requires
handling of a 3D system matrix, demanding high computational power.
Second, in addition to the quantitative performance metrics such as PSNR and SSIM, for clinical acceptance, perceptual scores provided by
radiologists, as in~\mycite{sanaat2020projection, subtleRadiology}, can provide insights.
Third, although suDNN shows robustness to OOD data by producing qualitatively superior PET images even with uLD-PET, a recalibration mechanism may
benefit clinical interpretation towards quantitative imaging.
Finally, while the size of the dataset used in this paper is larger than those used in the publications involving the baseline methods (M1--M5), we
plan to evaluate the proposed method on multiple cohorts, including covering healthy and pathological conditions.

In summary, our suDNN framework, informed by the underlying imaging physics and that models uncertainty/heteroscedasticity, achieves a more robust
mapping from uLD PET images (including the multi-contrast MRI) to SD-PET images. suDNN demonstrates robustness to unseen OOD PET acquisitions and
provides an estimate of the underlying uncertainty of the prediction, which facilitates a new paradigm of risk assessment in the application of DNNs
to low dose PET image reconstruction. The method has the potential to dramatically improve the utility of uLD PET imaging in diagnostic imaging,
therapeutic monitoring, and drug development research in oncology, neurology, and cardiology. Physics-inspired DNN-based reconstruction of low-dose
PET scans \revise{has} the potential to substantially expand the use of PET in longitudinal studies and imaging of radiation-sensitive populations, including
children and pregnant women.

\section*{Acknowledgment}

The authors are grateful for support from the Infrastructure Facility for Advanced Research and Education in Diagnostics grant funded by Department of
Biotechnology, Government of India (BT/INF/22/SP23026/2017). The authors would also like to acknowledge Siemens Healthineers for providing e7tools
software for PET image reconstruction. The authors thank Shenpeng Li (Monash University) for providing reconstructed PET images from scanner data.

\section*{References}
\bibliography{pet_ld2sd1}
\end{document}